\documentclass{iopjournal}

\usepackage{hyperref}
\usepackage{mathtools}
\usepackage{amsmath}
\usepackage{amssymb}
\usepackage{amsbsy}
\usepackage{bm}
\usepackage{textcomp}
\usepackage{xcolor}
\usepackage{fancyhdr}
\usepackage{float}
\usepackage{dsfont}
\usepackage[normalem]{ulem}
\usepackage{enumitem}

\newcommand{\Nmax}{$N_{\rm max}$}

\newcommand{\SU}[1]{\ensuremath{\mathrm{SU}( #1 )}}

\newcommand{\SO}[1]{\ensuremath{\mathrm{SO}( #1 )}}

\newcommand{\SpR}[1]{\ensuremath{\mathrm{Sp}( #1,\mathbb{R} )}}

\newcommand{\hw}{\ensuremath{\hbar\Omega}}

\newcommand{\NNLOopt}{NNLO$_\mathrm{opt}$}

\newcommand {\red} [1]{\textcolor{black}{#1}}
\newcommand {\IR} [1]{\textcolor{black}{#1}}

\newcommand{\bea}{\begin{eqnarray}}
\newcommand{\eea}{\end{eqnarray}}
\newcommand{\bitz}{\begin{itemize}}
\newcommand{\eitz}{\end{itemize}}

\begin{document}

\articletype{Paper} 

\title{\emph{Ab initio} optical potentials for magnesium isotopes \red{at intermediate energies}: from stability to the island of inversion}

\author{G. H. Sargsyan $^{1,*}$\orcid{0000-0002-3589-2315}, J. I. Fuentealba Bustamente $^2$\orcid{0009-0000-3472-4347}, K. Beyer $^1$\orcid{0000-0002-2695-2478} and Ch. Elster $^{2}$\orcid{0000-0002-2459-1226}}

\affil{$^1$Facility for Rare Isotope Beams (FRIB), Michigan State University, East Lansing, Michigan 48824, USA}

\affil{$^2$Institute of Nuclear and Particle Physics, and Department of Physics and Astronomy, Ohio University, Athens, OH 45701, USA}

\affil{$^*$Author to whom any correspondence should be addressed.}

\email{sargsyan at frib}


\begin{abstract}
We present the first calculations of \emph{ab initio} nonlocal optical potentials for $^{24,26,28}$Mg and $^{32}$Mg isotopes using the leading-order term of the spectator expansion of multiple-scattering theory. We use the structure input from the \emph{ab initio} symmetry-adapted no-core shell model (SA-NCSM), which provides translationally invariant, off-shell scalar and spin-projected densities so that structure and reaction inputs are treated on equal footing with no adjustable parameters. This leading-order potential reproduces $^{24}$Mg neutron total, reaction, and elastic-scattering data at energies between 65 and 250 MeV and provides predictions for $^{26,28}$Mg and $^{32}$Mg. We compare our predictions with those from the uncertainty-quantified Koning-Delaroche (KDUQ) and Weppner-Penney global optical potentials, as well as the ENDF nuclear data evaluations. These
comparisons highlight some of the limitations of the global models, while also validating their use in reaction modeling near the N=20 island of inversion.
\end{abstract}

\section{Introduction}
\label{sec:intro}
The study of nuclei often relies on nuclear reactions to extract observables related to structure and dynamics.
Reliably modeling the interaction between reaction fragments is key to extracting this information, especially as one moves toward the drip line. Oftentimes, for nucleon-nucleus reactions, one relies on global phenomenological optical potentials, which are fit to elastic cross sections on many stable targets, with some parametric dependence on target mass $A$ and charge $Z$, e.g., \cite{Koning:2003zz} (for a more comprehensive review, see \cite{Hebborn:2022vzm}). These phenomenological interactions are then extrapolated away from the valley of $\beta$-stability to describe reactions involving rare isotopes. Accordingly, quantifying the uncertainties of such extrapolations has become an important goal, leading to uncertainty-quantified global optical potentials such as KDUQ \cite{Pruitt_KDUQ2023}. It has been shown that, even when only considering reactions with nuclei in the valley of stability, there is a considerable uncertainty in reaction observables resulting from calibrating these phenomenological models to data ~\cite{Lovell:2016nps}, not to mention the uncertainty associated with assuming some simplified phenomenological form. Therefore, efforts to achieve reliable predictive power for reactions involving rare isotopes must also involve input from microscopic structure and reaction theory. Especially powerful within this category are {\it ab initio} techniques, capable of predicting reaction observables using nucleon-nucleon interactions consistent with quantum chromo-dynamics as the only inputs \cite{Hergert20,Ekstrom:2022yea}.

With advances in high-performance computing, significant progress has recently been achieved in \emph{ab initio} many-body methods for scattering and nuclear reactions (see reviews in Refs. \cite{ 0954-3899-41-12-123002,Navratil:2016ycn,Johnson:2019sps, Launey:2025qdd}). These methods have been applied to elastic scattering \cite{NollettPWCH07,HagenDHP07,Quaglioni:2008sm, QuaglioniN09,ElhatisariLRE15, Viviani:2020, MercenneLDEQSD21}, photoabsorption \cite{PhysRevC.90.064619}, transfer \cite{NavratilQ12}, capture \cite{PhysRevLett.105.232502}, thermonuclear fusion \cite{HupinQN19}, and alpha-capture processes  \cite{DreyfussLESBDD20}. Most of these studies, however, have been performed for very light nuclei, as solving the full many-body system, including continuum degrees of freedom, from first principles often becomes computationally intractable. For heavier nuclei, a more practical strategy is to isolate a few relevant degrees of freedom--typically the reaction fragments or clusters--and map the full many-body problem onto a few-body framework. This reduction produces effective interactions between clusters commonly called optical potentials. There is a growing demand for parameter-free interactions when studying nuclei far from stability, where elastic-scattering data alone cannot uniquely determine the optical potential, and uncertainties become large \cite{LovellN15}.

Depending on the energy regime under consideration, there are \red{only a few main} approaches to obtaining optical potentials from first principles. At low projectile energies ($\lesssim$ 20~MeV per nucleon), {\it ab initio} nucleon-nucleus ($NA$) effective interactions for elastic scattering are derived with Green's function techniques utilizing different many-body methods like the symmetry-adapted no-core shell-model (SA-NCSM)~\cite{Launey:2021sua, BurrowsLMBSDL24}, or the coupled cluster approach~\cite{RotureauDHJN2018}, with some recent methods allowing for use of microscopic structure input from any many-body model \cite{Sargsyan:2024qeb}. The goal of these approaches is to expand the dynamical part of the optical potential to include the low-energy collective degrees of freedom of the projectile-target composite system. \red{Another method rooted in nucleon-nucleon ($NN$) interactions derives optical potentials from chiral forces in nuclear matter \cite{WhiteheadLH2019,WhiteheadLH2020,WhiteheadLH2021}. These potentials produce proton and neutron elastic scattering cross sections and can be used for studies of reactions involving a single nucleon (see, e.g., \cite{WhiteheadPNP2022}).}

The present study concentrates on the energy regime between $\sim$60 and \red{250}~MeV per nucleon, for which the spectator expansion of multiple scattering~\cite{Siciliano:1977zz} \red{provides a well-established description. The same formalism can be applied at even higher energies, see, e.g., Refs.~\cite{Chinn:1993zza,Elster:1996xh}, where the leading order clearly dominates. In our present consideration, the limit for the highest energies under consideration is given by the chiral $NN$ potential we employ.}
Here, nuclear densities obtained within an {\it ab initio} framework are used consistently, with the same $NN$ interaction, in both structure and reaction calculations.  The pioneering work in deriving an {\it ab initio} effective interaction for $NA$ elastic scattering for intermediate projectile energies was based
on the no-core shell model (NCSM) and thus limited to light nuclei with masses up to
$A\simeq16$~\cite{Burrows:2018ggt,Burrows:2020qvu,Baker:2023uzx,Gennari:2017yez} for reasonably well-converged
calculations of nuclear densities. The SA-NCSM can push the structure calculations to higher mass nuclei ($A \simeq$
48~\cite{LauneyMD_ARNPS21,Burrows:2023ugy, Vorabbi:2023mml}) by considering
shape-related symmetries to construct the basis and select only non-negligible configurations.
The advantage of this selection process is
a drastic reduction in the number of basis states, which in turn allows calculations to move toward heavier nuclei.

In this work, the non-local, translationally invariant scalar one-body densities and  spin-projected momentum
distributions  are derived from the SA-NCSM and employed for the calculation of a leading-order -- in the spectator
expansion --  effective
$NA$ interaction for targets of nuclei in the magnesium isotopic chain. 
In particular, we study $^{24, 26, 28}$Mg and $^{32}$Mg, which span from the stable $N=Z$ to the short-lived $N=20$ island of inversion region. The study of these isotopes is especially compelling because all of them exhibit deformation and strong collectivity \cite{Martin:2024}, and thus the features of the SA-NCSM are well-suited to compute the translationally invariant one-body density matrix within this scheme. In addition, it is found in Ref.~\cite{Burrows:2020qvu} for $^6$He and $^8$He and in Ref.~\cite{Baker:2024wtn} for $^{20}$Ne that the {\it ab initio} leading order of the spectator expansion describes scattering observables slightly better for halo or deformed open-shell nuclei with a $0^+$ ground state than for spherical, closed-shell nuclei as e.g. $^{40}$Ca~\cite{Baker:2024wtn}. A reason may be that, for open-shell deformed nuclei, rescattering of a projectile within the nucleus is less important. Therefore, we first study $^{24}$Mg, for which experimental information is available in the energy range 65-250~MeV per nucleon, to confront the theoretical prediction with experiment. Then we will move along the Magnesium isotope chain and predict scattering observables in that energy regime. \IR{We will also} compare with predictions of phenomenological optical potentials
to investigate how extrapolated phenomenology compares with our theoretical predictions. \IR{For this comparison} we concentrate on the predictions from the  KDUQ~\cite{Pruitt_KDUQ2023,bandframework} optical potential, \IR{which is based on the Koning-Delaroche  (KD) optical potential~\cite{Koning:2003zz}  and additionally} includes uncertainty quantifications. \IR{Since the KD potential is widely used in standard reaction codes, we considered it an appropriate choice.}

The structure of the paper is as follows. In Section \ref{sec:theory}, we briefly discuss the theoretical methods behind the many-body \emph{ab initio} model that we use in this work, the leading-order effective nucleon-nucleus potential, and the global optical potentials commonly used by the nuclear physics community. Section \ref{sec:results} contains the results of the scattering calculations and their discussions. In particular, Section \ref{sec:24Mg} presents our calculations of neutron and proton scattering observables on $^{24}$Mg, compared with experimental data and global optical potentials. Section \ref{sec:Mg_rest} contains our predictions for $^{26, 28}$Mg and $^{32}$Mg with comparisons to the global models where applicable. We summarize and offer perspectives on improving our leading-order effective potential and the global optical model parametrizations in Section \ref{sec:summary}.

\section{Theoretical framework}
\label{sec:theory}

\subsection{Symmetry-adapted no-core shell model}

For the calculation of nuclear one-body densities, we utilize the \textit{ab initio} symmetry-adapted no-core shell model (SA-NCSM) \cite{LauneyDD16,DytrychLDRWRBB20,LauneyMD_ARNPS21}. This model treats all nucleons on equal footing and allows for chiral EFT interactions among them. The incorporation of chiral EFT interactions aligns with the chiral symmetry and symmetry-breaking patterns inherent to quantum chromodynamics, thereby facilitating nuclear calculations based solely on the dynamics of two- and three-nucleon systems. Additionally, the symmetry-adapted (SA) basis within the many-body SA-NCSM framework offers solutions for otherwise intractable ultra-large model spaces~\cite{LauneyDD16, DytrychLDRWRBB20}. These solutions are crucial for accurately depicting complex nuclear properties, such as clustering, collective behavior, and interactions with the continuum.

Similar to the no-core shell model (NCSM) \cite{NavratilVB00,NavratilVB00b}, SA-NCSM uses a harmonic oscillator (HO) basis, where the HO major shells are separated by a parameter
$\hbar \Omega$. 
The size of the model space is given by an \Nmax~cutoff, which is the maximum total number of oscillator quanta above the lowest HO configuration for a given nucleus. The SA-NCSM calculations use a non-relativistic nuclear Hamiltonian with translationally invariant interactions plus Coulomb interaction. The model computes the eigenvalues and eigenvectors of the nuclear interaction Hamiltonian, using the eigenvectors to evaluate nuclear observables. As \Nmax~increases, the results converge to the exact value, becoming independent of the HO parameter $\hbar \Omega$ in the limit as \Nmax $\rightarrow \infty$ (comprising infinitely many HO shells). It is important to emphasize that the results obtained from SA-NCSM exactly match those from the conventional NCSM \cite{NavratilVB00,Barrettnv13} when using the same nuclear interaction and the model space. However, SA-NCSM leverages the emergent symplectic symmetry \SpR{3}$\supset$\SU{3}$\supset$\SO{3} in nuclear systems \cite{DytrychLDRWRBB20}, where each \SpR{3}-invariant subset of basis states characterizes a specific nuclear shape, and each \SU{3}-invariant subset informs particular deformations. This approach allows the SA-NCSM to achieve computational convergence and effectively describe clustering and deformation in nuclei using only a fraction of the relevant NCSM space~\cite{DreyfussLESBDD20,SargsyanLBGS2022}.

Within the SA-NCSM selected model spaces, the spurious center-of-mass motion can be exactly factored out from the intrinsic dynamics~\cite{Verhaar60,Hecht71} (see, e.g., Ref. \cite{Burrows:2023ygq}). This plays a vital role in calculations of scattering observables since the necessary one-body densities computed in the SA-NCSM are exactly translationally invariant (without any center-of-mass spuriosity). 


In this work, the fully microscopic wavefunctions for the ground states of $^{24}$Mg, $^{26}$Mg, $^{28}$Mg, and $^{32}$Mg are computed using the NNLO$_{\rm opt}$ $NN$ chiral potential \cite{Ekstrom13} within the \textit{ab initio} SA-NCSM many-body framework. This potential has produced energy spectra and observables (such as radii, quadrupole moments, $E2$ transitions, and charge form factors) that are in close agreement with experimental results \cite{LauneyDD16, DytrychLDRWRBB20,LauneyMD_ARNPS21}. For the calculations of $^{24}$Mg, $^{26}$Mg, $^{28}$Mg, we have used \Nmax=4, 6, and 8 model spaces and \hw=13, 15, and 17 MeV, which have been found to be optimal for the convergence of reaction observables in this mass region. For $^{32}$Mg we present calculations in \Nmax=6 model space and \hw=15 MeV only because of the high computational cost. Except for the \Nmax=4 model space of $^{24}$Mg, all other model spaces have been selected from a subset of the NCSM space using the SA basis following the procedure described in Refs. \cite{Dytrych:2016cc,Baker:2024wtn}.

The NNLO$_{\rm opt}$ potential \cite{Ekstrom13} is calibrated exclusively to $NN$ scattering phase shifts up to 135 MeV and deuteron properties, without utilizing three-nucleon forces, which have been shown to have a negligible effect on the binding energies of three- and four-nucleon systems in calculations with this interaction \cite{Ekstrom13}. Moreover, the NNLO$_{\rm opt}$ $NN$ potential has demonstrated the ability to reproduce a range of observables, providing results similar to those derived from chiral potentials that incorporate three-nucleon forces. These observables include the electric dipole polarizability of $^4$He \cite{BakerLBND20}, the analyzing power for elastic proton scattering on $^4$He, $^{12}$C, and $^{16}$O \cite{BurrowsEWLMNP19}, as well as B(E2) transition strengths for $^{21,28}$Mg and $^{21}$F \cite{Ruotsalainen19,PhysRevC.100.014322} in the SA-NCSM without effective charges. Additionally, it encompasses collective and dynamical observables for $^{20}$Ne \cite{SargsyanYOLELT25},  $^{40}$Ca \cite{Baker:2024wtn,Burrows_2025} and $^{48}$Ti \cite{LauneyMD_ARNPS21}. From this point of view, the NNLO$_{\rm opt}$ chiral $NN$ interaction is very well suited for elastic scattering calculations using an optical potential based on the leading order in the spectator expansion, since this order only contains explicit two-nucleon forces.

\subsection{Leading-order {\it ab initio} effective nucleon-nucleus potential}

Calculating elastic nucleon-nucleus ($NA$) scattering observables in an {\it ab initio} fashion requires not only the interactions between the nucleons in the target but also the interaction between the projectile and the nucleons in the target. A multiple-scattering expansion provides a framework for organizing this process in a tractable way. The spectator expansion~\cite{Siciliano:1977zz,Baker:2023uzx} organizes the scattering of a nucleon from a nucleus consisting of $A$ nucleons in terms of nucleons participating actively in the scattering process. In the leading order of the spectator expansion, there are two active nucleons involved: the projectile and a target nucleon. The next-to-leading order would involve the projectile and two target nucleons, and so on. Thus, by construction, the leading order term of the spectator expansion requires the two-nucleon force between the projectile and a target nucleon. 
A scalar one-body density and a spin-projected momentum distribution, both nonlocal and translationally invariant, represent the struck nucleon in the target. Those density distributions are then folded with off-shell $NN$ amplitudes given in the Wolfenstein parameterization~\cite{wolfenstein-ashkin,Wolfenstein:1956xg}. To ensure that the $NN$ interaction is treated consistently in the structure and reaction parts of the calculation, folding with both the scalar one-body density and the spin-projected momentum distribution is necessary. We refer interested readers to Ref.~\cite{Burrows:2020qvu} for the formal derivation of the leading order $NA$ effective interaction, and to Ref.~\cite{Baker:2024wtn} for explicit expressions.
In this work we concentrate on the Magnesium isotopes with $J^\pi= 0^+$ and construct the effective interaction for e.g. proton scattering as $\widehat{U}_{\mathrm{p}}(\bm{q},\bm{\mathcal{K}}_{NA};\epsilon)$, which is a function of the momentum transfer ${\bf q}$ and the average momentum ${\bm{\mathcal{K}}_{NA}}$, where the
subscript $NA$ refers to the nucleon-nucleus frame, as well as the energy $\epsilon$, which in the impulse approximation is given by the corresponding energy of the $NN$ amplitudes.  For the explicit expression and further explanations, we refer the reader to Ref.~\cite{Baker:2024wtn}.

When calculating $NA$ elastic scattering amplitudes,
the leading order term in the spectator expansion, $\widehat{U}_{\mathrm{p}}$, does not directly enter a Lippmann-Schwinger type integral
equation for the transition amplitude for $NA$ scattering. \IR{To obtain the correct Watson optical potential $U_p(\bm{q},\bm{\mathcal{K}}_{NA}; \epsilon)$ that enters the $NA$ scattering equation, an additional step needs to be carried out. 
According to Ref.~\cite{Baker:2023uzx} the term $\widehat{U}_{\mathrm{p}}$ is comprised of a sum over proton and neutron densities folded with the corresponding proton-neutron (pn) and proton-proton (pp) Wolfenstein amplitudes. For each term an additional integral equation is solved,
\begin{equation}
\label{eq:watson}
U_{\rm p \alpha} = \widehat{U}_{\rm p \alpha} - \widehat{U}_{\rm p \alpha} G_0 (E) P U_{\rm p \alpha},
\end{equation}
where $\alpha$ indicates either proton or neutron contributions. The Watson optical potential is then given by the two contributions. A detailed derivation is given in Ref.~\cite{Chinn:1993zza}. }
For simplicity, the momentum variables are omitted in Eq.~(\ref{eq:watson}).
The free $NA$ propagator is give by $G_0(E)$ and $P$ is a projector on the ground state. As pointed out in Ref.~\cite{Baker:2023uzx}, for solving the scattering problem, the reference energy separating bound and continuum states is chosen such that the ground state energy is set to zero, implying that the energies referring to the target Hamiltonian in $G_0 (E)$ are excitation energies of the target. The Watson optical potential is required to properly account for the difference between proton and neutron densities \IR{folded with} the corresponding proton-proton and neutron-proton $NN$ amplitudes. This becomes important when predicting scattering observables for nuclei closer to the dripline, where the neutron and proton numbers differ.
For proton-nucleus scattering calculations, the Coulomb interaction between the projectile and the target is included using the exact momentum space formulation described in Ref.~\cite{Chinn:1991jb}.

\subsection{Phenomenological optical potentials}

Computing {\it ab initio} effective interactions in leading order of the spectator expansion for exotic nuclei gives a unique opportunity to compare to phenomenological optical potentials that are fitted within the valley of stability and then extrapolated to exotic nuclei.
In this work, we concentrate on the Koning-Delaroche Uncertainty Quantified (KDUQ) global optical potential \cite{Pruitt_KDUQ2023}. This model takes the phenomenological form of the original Koning-Delaroche global potential \cite{Koning:2003zz}, and recalibrates its 47 free parameters using a full Bayesian approach. Uncertainty quantification is achieved by drawing parameter samples from the KDUQ posterior and propagating them through to cross sections. In this work, we compare to the federal KDUQ posterior; the version of the posterior in which each independent experiment is equally weighted in the likelihood, using the 416 parameter samples provided in the supplemental material of \cite{Pruitt_KDUQ2023}. We use jit$\mathcal{R}$, the just-in-time $\mathcal{R}$-matrix \cite{Beyer_JITR_2024}, which includes a built-in implementation of KDUQ, for phenomenological reaction calculations.

The phenomenological form used by KDUQ is energy-dependent and complex. Unlike the \textit{ab initio} approach, it is local in coordinate space. Its radial dependence includes a central part composed of real and complex Woods-Saxon-based volume terms and a complex surface-peaked term, as well as a spin-orbit part composed of real and complex Thomas forms. Each of these terms has a fixed diffuseness and a radius that depends only on $A^{1/3}$. The depths of each of these terms are parameterized to depend on energy, $A$, and $(N-Z)/A$, with the complex surface term dominant at low energies $E \lesssim 50$ MeV to describe absorption at the nuclear surface through excitation of collective degrees of freedom, and the complex volume term dominant for $E \gtrsim 50$ MeV, to describe absorption in the interior. Some of the free parameters associated with this phenomenological form are fit separately for neutrons and protons, so it is not Lane consistent \cite{LANE1962676}. 

Experimental data included in the KDUQ calibration ranged from laboratory kinetic energies of 1 keV to 200 MeV, for (near-)spherical, stable nuclei with masses $A=24$ to 209. Therefore, application to the Mg chain represents the lower extreme of its expected region of applicability in mass, and extrapolation beyond stability and target sphericity is beyond its limits. Specific to the Mg chain, the KDUQ corpus includes elastic neutron scattering on $^{24}$Mg and $^{\text{nat}}$Mg up to about 20 MeV, and total neutron cross sections up to about 50 MeV, but no data for proton projectiles. Therefore, one would expect its extrapolation beyond stability in the Mg chain to be under-constrained, making it an interesting comparison with our \textit{ab initio} approach.

For neutron total cross sections and proton total reaction cross sections, we also compare to the suggestions given by the current ENDF nuclear data evaluation~\cite{ENDF/B-VI} and the iso-spin dependent global nucleon-nucleus phenomenological optical potential by Weppner, Penney et al.~\cite{Weppner:2009qy}. The Weppner-Penney (WP) phenomenological optical potentials are energy-dependent, based on Woods-Saxon terms, and are fitted to the 2009 experimental database. \IR{However, it does neither offer uncertainty quantification nor can it be extrapolated too far from the valley of stability.} Therefore, we do not use it in comparison of differential scattering observables.

\section{Results and discussion}
\label{sec:results}

\subsection{Scattering Observables for $^{24}$Mg}  
\label{sec:24Mg}
\begin{figure}[t]
\centering
\includegraphics[width=9cm]{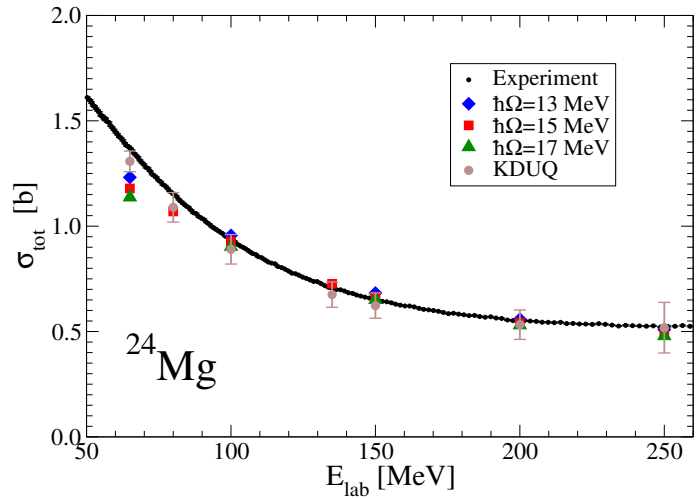}
\caption{The total cross section for neutron scattering from $^{24}$Mg as a function of the laboratory kinetic energy for N$_{\rm max}$=6  for the \hw~values 13, 15, and 17 MeV together with the predictions of  KDUQ \cite{Pruitt_KDUQ2023}. 
The experimental data are from 
Ref.~\cite{Abfalterer:2001gw}. 
}
\label{fig:Mg24-ntot-hw15-exp}
\end{figure}

The starting point for our study of elastic scattering observables from Magnesium isotopes is $^{24}$Mg, which consists of an equal amount of protons and neutrons. However, $^{24}$Mg is deformed and thus a good candidate to test how well predictions at leading order in the spectator expansion for elastic scattering observables hold. As the first observable, we consider the total cross section for neutron scattering from $^{24}$Mg, which has been measured accurately over a wide energy range~\cite{Abfalterer:2001gw}.
The predictions of the leading-order spectator expansion based on the NNLO$_{\rm opt}$ chiral interaction and SA-NCSM one-body densities are shown in Fig.~\ref{fig:Mg24-ntot-hw15-exp} for energies between 65 and 250~MeV. The calculations are carried out for three different values of \hw~(13, 15, and 17 MeV). For laboratory kinetic energies between 100 and 250 MeV, the {\it ab initio} prediction provides an excellent description of the experiment. A slight deterioration of the agreement is observed at 65 and 80 MeV, which is, however, not surprising, since with decreasing energy, the probability of rescattering within the nucleus increases, and the leading-order in the spectator expansion becomes insufficient. A qualitatively similar observation was made in \cite{Chinn:1994dh}.
The figure also shows that the dependence on the $\hw$-grids is almost negligible above 100 MeV scattering energy and quite small below. We tested the dependence on the size of the model space, $N_{\mathrm{max}}$, and found it to be even smaller. In addition, the predictions of the KDUQ phenomenological potential are also in very good agreement with the experimental values, albeit with uncertainties increasing at higher energies (Fig. \ref{fig:Mg24-ntot-hw15-exp}). 
\red{Since the applicability range of KDUQ is up to 200 MeV, the calculation at 250 MeV serves as an extrapolation outside of this range.} We also note that neutron scattering experimental data on $^{24}$Mg was used in the fits of KDUQ \cite{Koning:2003zz, Pruitt_KDUQ2023}.
\begin{figure}[h]
\centering
\includegraphics[width=7.5cm]{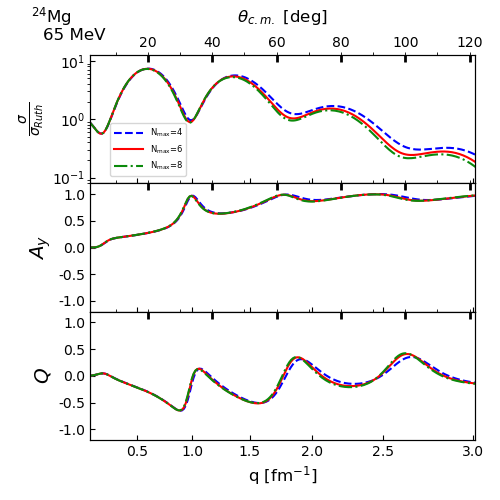}
\includegraphics[width=7.5cm]{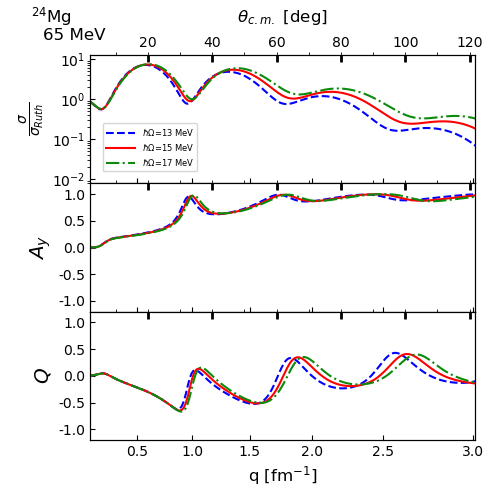}
\caption{The differential cross section, the analyzing power $A_y$, the spin rotation function $Q$ as a function of the momentum transfer $q$ (and c.m. angle $\theta_{c.m.}$) for proton scattering from $^{24}$Mg at 65 MeV laboratory kinetic energies as function of  N$_{\rm max}$ (left) for $\hbar\Omega$=15 MeV and $\hbar\Omega$ for \Nmax=6 (right). 
}
\label{fig:Mg24a-hw15-exp}
\end{figure}

To learn about the dependence of scattering observables on the SA-NCSM model space parameters \Nmax~and \hw,
we therefore should consider angle (momentum) dependent observables, i.e.,
 proton elastic differential scattering cross sections, analyzing powers, and spin rotation functions over the variation of \Nmax~for a fixed \hw, and variation of \hw~for a fixed \Nmax~(Fig. \ref{fig:Mg24a-hw15-exp}). Varying the model space between \Nmax=4 and 8, we see \IR{very little} change for all the scattering observables throughout the entire angular range (Fig. \ref{fig:Mg24a-hw15-exp}a), which indicates that the chosen \hw~values are optimal. This is consistent with earlier studies of lighter systems \cite{Burrows:2018ggt, Baker:2024wtn}. \red{We note that, for an optimal \hw, the uncertainty because of the selection of the model space in SA-NCSM is expected to be smaller than variation in \Nmax, as also shown in the study of Ref. \cite{Baker:2024wtn}.} On the other hand, examining the \hw~variance for \Nmax=6 reveals noticeable change in larger angles, especially for the cross sections (Fig. \ref{fig:Mg24a-hw15-exp}b). We observe the same trends across all the isotopes and observables discussed in this paper. Hence, for the rest of the manuscript, we present results for \Nmax=6 with \red{this} \hw~variance as uncertainties \red{in our calculations. This should not be confused with a complete estimate of chiral uncertainties, as, e.g., carried out in Ref.~\cite{Baker:2023uzx}.}

\begin{figure}[h]
\centering
\includegraphics[width=7.5cm]{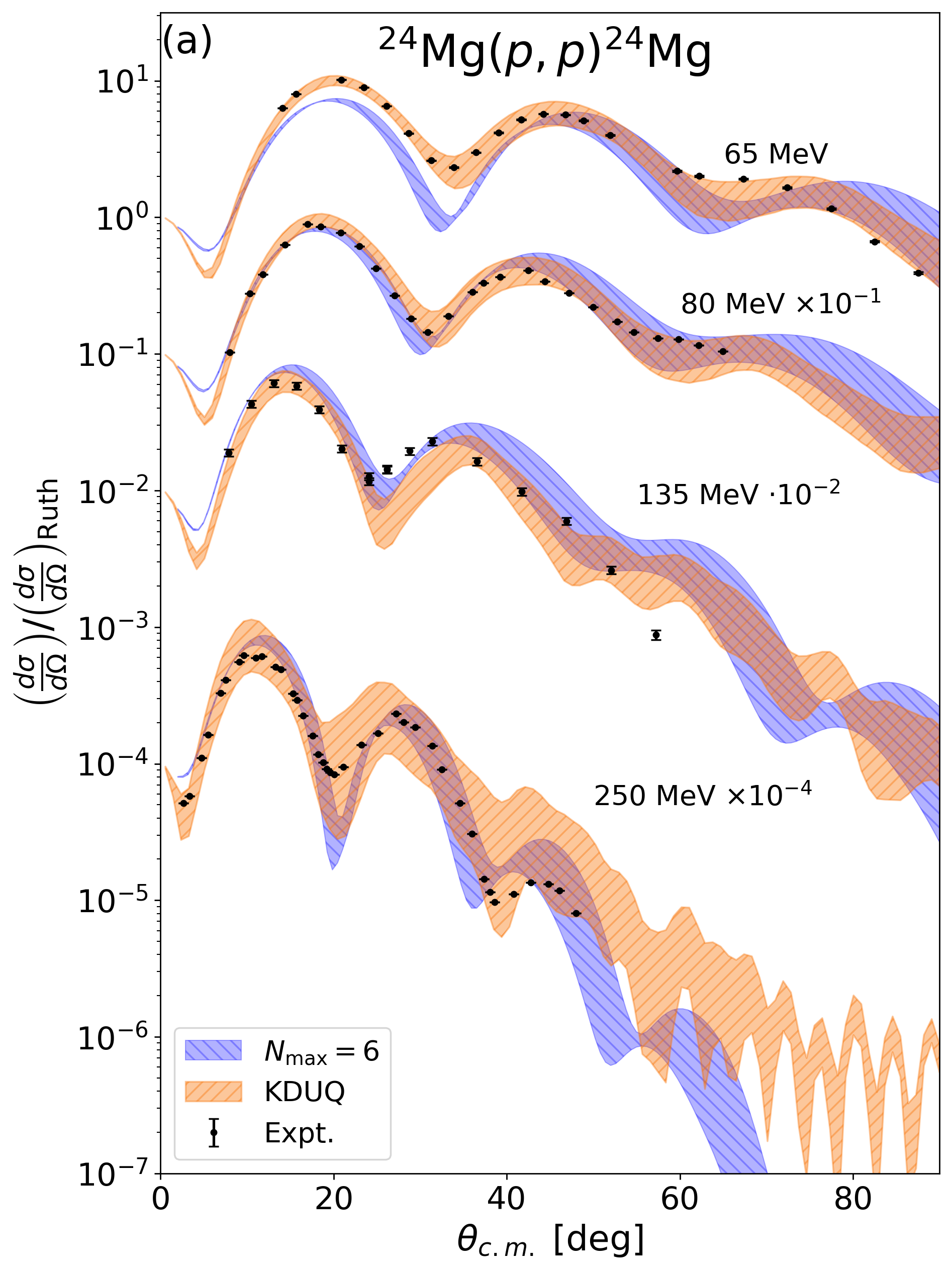}
\includegraphics[width=7.2cm]{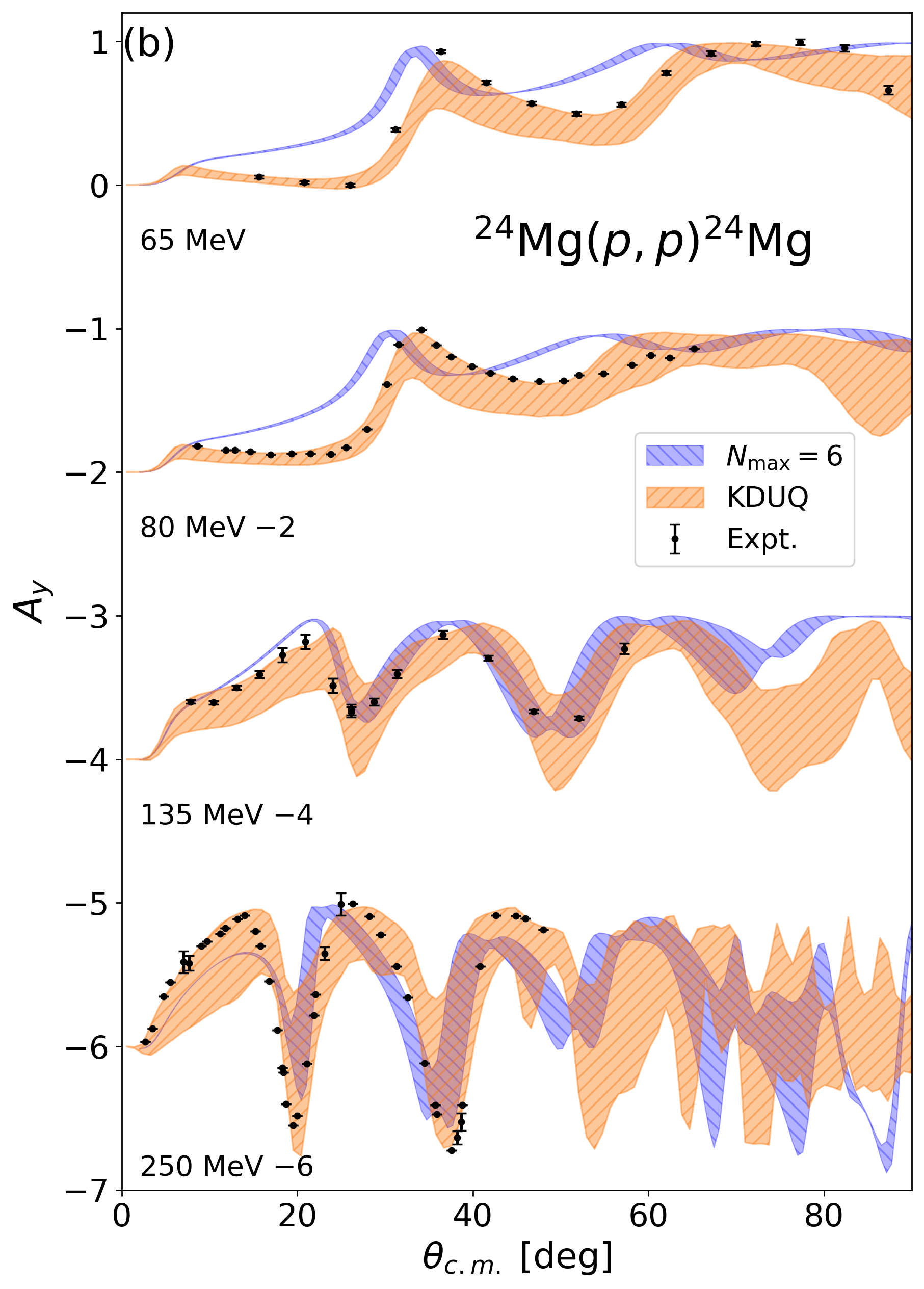}
\caption{The angular distribution of the differential cross section divided by the Rutherford cross section (a) and analyzing power (b) as a function of the c.m. scattering angle for elastic proton scattering from $^{24}$Mg from 65 to 250 MeV laboratory projectile energy. The band on ``\Nmax=6'' calculations represents the variation of $\hw$ between 13 and 17 MeV. The cross sections are multiplied by the powers of 10 indicated at the energies listed in the figure, while the analyzing powers are offset by -2. The data at 65 MeV are from \cite{Kato:1985zz,Sakaguchi:1979fpk}, the data at 80 MeV from \cite{Hatanaka:1984zz}, the data at 135 MeV from \cite{Schwandt:1982py}, and those at 250 MeV from \cite{Hicks:1988zzb}.}
\label{fig:24Mg_stacked}
\end{figure}

Next, we 
study the energy dependence of the elastic scattering observables for proton scattering from $^{24}$Mg. 
The calculated differential cross sections and analyzing powers are shown in Fig.~\ref{fig:24Mg_stacked} for laboratory kinetic energies between 65 and 250 MeV in comparison with available experimental data. The variation in $\hw$ between 13 and 17 MeV in the calculations is indicated by the left-hatched (purple) error band. In general, our {\it ab initio} predictions of the scattering observables are in very good agreement with the experimental data.
Smaller scattering angles correspond to lower momentum transfer. The narrow \hw~bands at small angles indicate that the optical potential probes the low-$q$ part of the nuclear density that is less sensitive to the \hw~variance (see Appendix A). Similar to the case of neutron cross sections, the slight deviations from the experimental data for 65 MeV result are most likely from the rescattering effects not included at leading order, while deviations at 250 MeV might \red{reflect} the fact that the chiral NNLO$_\mathrm{opt}$ interaction is \red{less well calibrated} to $NN$ data \red{at these higher} energies. 

In addition, we compare our calculated cross sections with those from KDUQ, represented by the right-hatched orange bands in Fig. \ref{fig:24Mg_stacked}. For small angles, the cross-section values and their minima from both calculations agree. However, noticeable discrepancies arise at higher angles (larger $q$ values). Notably, for the 250 MeV cross sections, the KDUQ values oscillate rapidly after 70 degrees, decreasing more slowly than our results.  Given that 250 MeV lies outside the energy range for which KDUQ is applicable, this finding also serves as a caution against extrapolating KDUQ beyond its recommended energy range.

Furthermore, for a more detailed study of the energy dependence of the elastic scattering observables, we calculate the analyzing power ($A_y$) of $^{24}$Mg($p,p$)$^{24}$Mg for the same energies as in elastic scattering cross sections ( Fig. \ref{fig:24Mg_stacked}(b)). Similar to the cross sections, the agreement with the experimental data improves at higher energies. Notably, for $\theta_{c.m.} < 25$ degrees, the experiment recorded virtually no analyzing power at 65 and 80 MeV, indicating no discernible difference between the elastic scattering cross sections of spin-up and spin-down polarized protons. In contrast, our calculated $A_y$ values increase gradually over this angular range. The calculations performed with KDUQ align well with the experimental data within their error bands; however, uncertainties tend to grow at higher energies, and at certain angles, they encompass nearly the entire range of the $A_y$ values. 

\begin{figure}[hbt]
\centering
  \centering
\includegraphics[width=7.2cm]{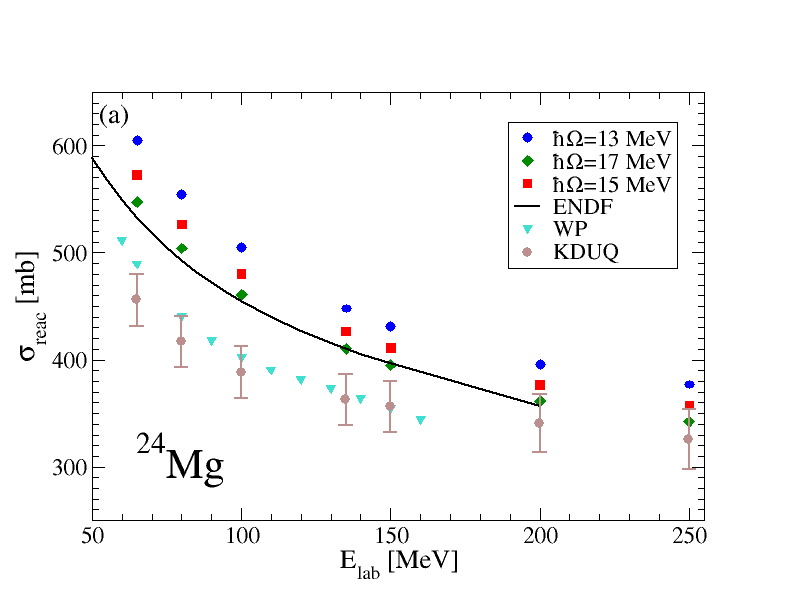}
\includegraphics[width=7.2cm]{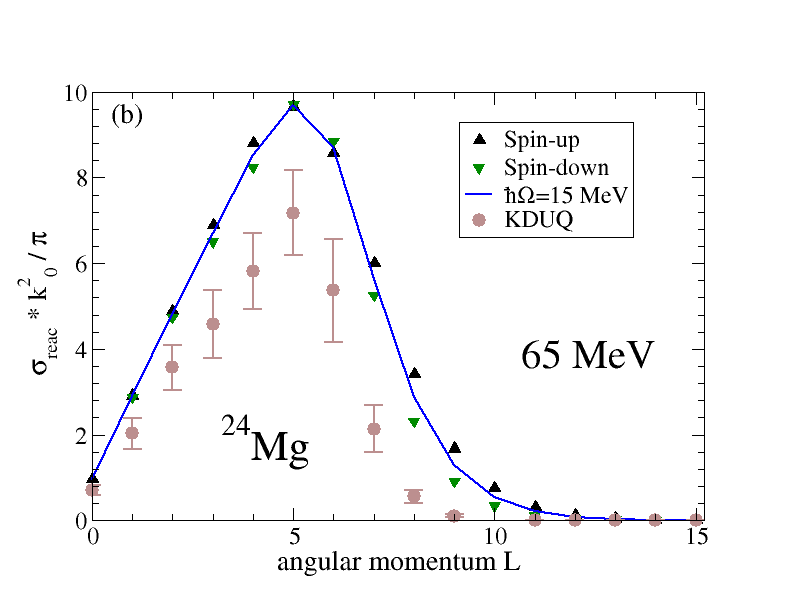}
  \caption{(a) The total reaction cross section for proton scattering from $^{24}$Mg as function of the laboratory kinetic energy for N$_\mathrm{max}$=6 and $\hw$=13, 15, and 17 MeV together with the ENDF~\cite{ENDF/B-VI}, Weppner-Penney~\cite{Weppner:2009qy}, and KDUQ~\cite{Pruitt_KDUQ2023} predictions. (b) The partial reaction cross sections multiplied by $k_0^2/\pi$ as a function of the angular momentum at 65 MeV laboratory kinetic energy, as calculated 
  for \Nmax=4 and $\hw$=15 MeV compared to the KDUQ prediction. See text for full explanation.
}
\label{fig:preaction-24Mg}
\end{figure}

Last, we consider the predictions for the proton total reaction cross section given in Fig.~\ref{fig:preaction-24Mg}. Here, we clearly see a stronger $\hw$ dependence than predicted for the neutron total cross section. That may be explained by the stronger $\hw$ dependence of the proton differential cross sections at larger angles (higher momentum transfers). We compare our results with those from the KDUQ and WP global optical potentials as well as the ENDF nuclear data evaluations \cite{ENDF/B-VI}. The two phenomenological optical potentials, KDUQ and WP, are consistently lower than our {\it ab initio} calculations, whereas the ENDF projection is closer to them. 
Considering the predicted differential cross sections from the KDUQ potential in Fig. \ref{fig:24Mg_stacked} may lead to a speculation that the overprediction of the differential cross section at larger angles may increase the integrated differential cross section, which in turn can influence the reaction cross section. Though there is no total cross-section for proton scattering, we saw earlier in Fig.~\ref{fig:Mg24-ntot-hw15-exp} that the neutron total cross-section depends only minimally on $\hw$. Since the reaction cross sections are calculated via the optical theorem, if the nuclear elastic cross section is larger, the reaction cross section will be smaller if the total is not affected.
As further illustration, we show in Fig.~\ref{fig:preaction-24Mg}(b) the partial reaction cross section as a function of angular momentum. \IR{The reaction cross section for a spin-$\frac{1}{2}$-spin-0 system is defined as
as
\begin{equation}
\sigma_{\mathrm reac}  =\frac{\pi}{k^2_0} \sum_L \left[ (L+1)(1-|S_{J-1/2}|^2) +L(1-|S_{J+1/2}|^2) \right],
\label{eq:reaction-cross-section}
\end{equation}
where $k_0$ is the c.m. momentum of the system  and $L$ is the orbital angular momentum. The partial waves are characterized by $L=J\pm 1/2$ partial waves.} Those are indicated by the upward and downward triangles in  Fig.~\ref{fig:preaction-24Mg}(b).
\IR{The calculated reaction cross section is depicted as blue line in Fig.~\ref{fig:preaction-24Mg}(b).}

\begin{figure}
    \centering
    \includegraphics[width=0.6\linewidth]{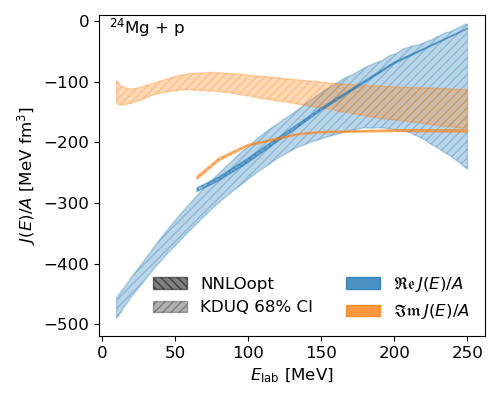}
    \caption{Comparison of the real (blue) and imaginary (orange) part of the volume integrals per nucleon for the central potentials of KDUQ (displaying 68\% credible intervals) and the local-equivalent extraction of the $\textit{ab initio}$ prediction for the calculation based on $\hbar\Omega$=13 to 17 MeV and \Nmax=6. See text for details.}
    \label{fig:mg24_volint}
\end{figure}

A sometimes useful intuitive metric for comparing optical potentials is the volume integral, which, for spherical potentials, can be written as 
\begin{equation}
\label{eq:volint}
    \frac{J(E)}{A} \equiv \frac{4\pi}{A} \int_0^\infty U_C(r;E) r^2 dr .
\end{equation}
For extracting an equivalent quantity from a nonlocal optical potential, we follow a procedure suggested in Ref.~\cite{Baker:2023wla} for obtaining the local part of a nonlocal potential 
$U_C(q,{\mathcal K},\theta_{q,{\mathcal K}};E)$, where
$\theta_{q,{\mathcal K}}$ is the angle between the vectors ${\bf q}$
and ${\bm {\mathcal K}}$. The potential surface for $\theta_{q,\mathcal{K}}=90^o$ is special, since here the
on-shell condition $q^2 + 4 \mathcal{K}^2 = 4 k_0^2$  is defined. 
With the help of this on-shell condition, we can obtain from the potential surface 
$U_C(q,\mathcal{K},\theta_{q,\mathcal{K}}=90^o;E)$ a function $U_C(q;E)$ at a given scattering energy, which then can be Fourier transformed to obtain $U_C{(\zeta)}$, where the radial variable $\zeta=\frac{1}{2}(r'+r)$ is the conjugate variable to the momentum transfer $q$. Thus, the equivalent of Eq.~(\ref{eq:volint}) for the {\it ab initio} potential is given by $\frac{4\pi}{A} U_C(q=0;E)$.

Of note are the systematic differences in absorbing power between KDUQ and our microscopic potential, with the latter yielding larger reaction cross sections across all energies. Additionally, when KDUQ is extrapolated to higher energies ($E > 200$ MeV), the total cross section does not decrease as rapidly as expected. 
Both of these trends can be understood by comparing volume integrals of KDUQ with our microscopic approach, as in Fig. \ref{fig:mg24_volint}. This figure shows that KDUQ has less imaginary strength than the microscopic approach, especially at low energies, and that for $E > 200$ MeV, some of the KDUQ samples reach a minimum in strength of the real potential and begin increasing again. Volume integrals for the other Mg isotopes show the same trend (see Appendix B). We also note that the bands on the KDUQ volume integrals represent sampling of around 40 parameters that are fitted to a large set of experimental data. The bands on our microscopic potential correspond to the \hw~variance only between 13 and 17 MeV in the calculations. The \hw~dependence is expected to further decrease with increasing model space. 

\begin{figure}[h]
\centering
\includegraphics[width=6.9cm]{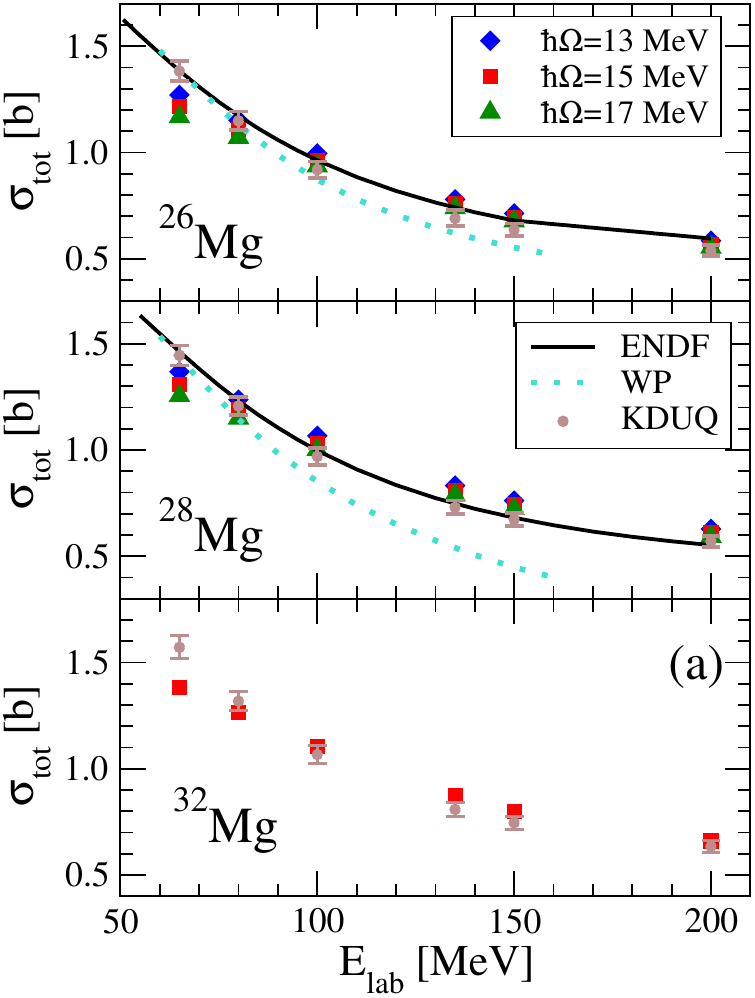}
\includegraphics[width=7.15cm]{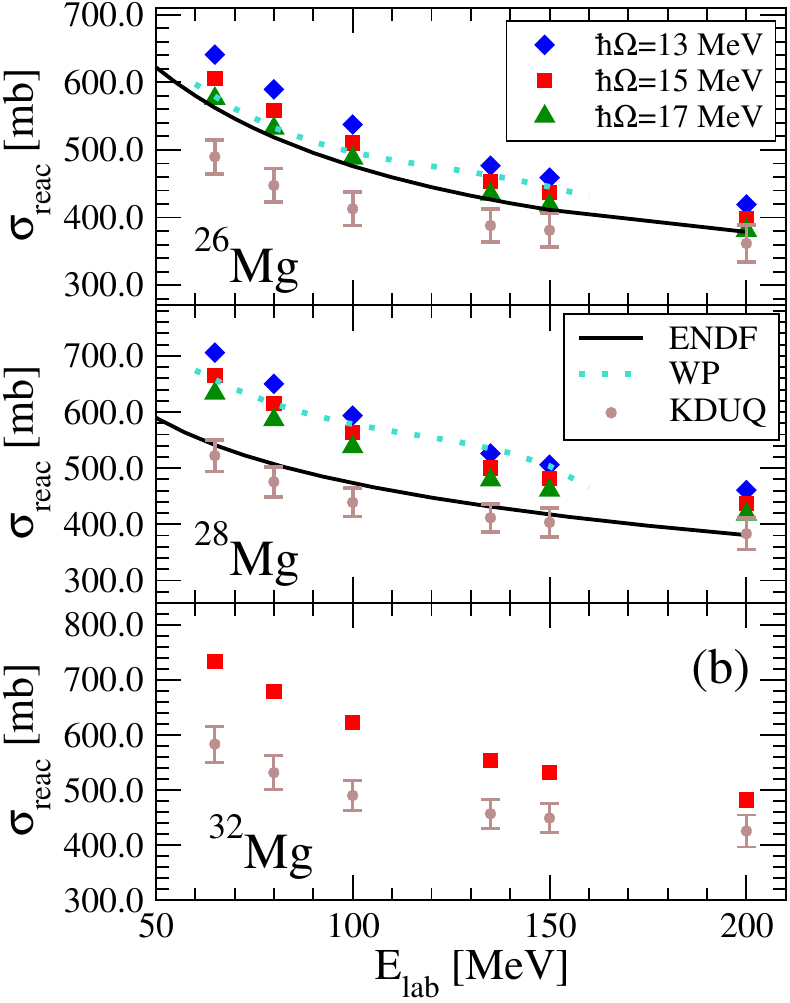}
\caption{The neutron total cross sections for $^{26}$Mg, $^{28}$Mg, and $^{32}$Mg as a function of the laboratory kinetic energy (left panel) and for the proton reaction cross sections (right panel). All {\it ab initio} calculations are performed with \Nmax=6 and the $\hw$ value indicated in the legend. For $^{26}$Mg and $^{28}$Mg, the ENDF projection and the WP phenomenological potential are included. For all isotopes, the KDUQ extrapolation is given.
}
\label{fig:26allMg_neut}
\end{figure} 

\subsection{Scattering observables for $^{26}$Mg, $^{28}$Mg, and $^{32}$Mg}
\label{sec:Mg_rest}
After comparing our calculations with the available experimental data for $^{24}$Mg, we present predictions for neutron total cross sections, as well as proton reaction and elastic scattering cross sections for target nuclei $^{26}$Mg, $^{28}$Mg, and $^{32}$Mg. Each of these isotopes exhibits deformation, with some having a mix of different dominant deformed configurations \cite{PhysRevC.100.014322}, hence posing significant challenges when modeled using microscopic nuclear structure methods. Notably, $^{32}$Mg resides in the $N=20$ island of inversion, where its ground state is anticipated to be predominantly shaped by configurations from the pf-shell. SA-NCSM \red{calculations have been shown to} successfully reproduce the low-lying excitation \red{spectra and the $B(E2)$ transition strengths of the neutron rich $^{28}$Mg and $^{32}$Mg isotopes} \cite{Launey:2025qdd}. \red{For $^{32}$Mg, SA-NCSM} calculations suggest a ground state structure characterized by a 2-particle-2-hole dominance. We use the densities of the ground states obtained from SA-NCSM \red{wavefunctions} (see Appendix A) to construct our microscopic optical potentials for $^{26}$Mg, $^{28}$Mg, and $^{32}$Mg isotopes. 

The neutron total cross sections for $^{26}$Mg, $^{28}$Mg, and $^{32}$Mg show a gradual increase from $^{24}$Mg, as illustrated in Fig. \ref{fig:26allMg_neut}(a). Consistent with the observations for $^{24}$Mg, our calculations reveal minimal dependence on the \hw~variation. \red{Since experimental data is not available} for $^{26}$Mg and $^{28}$Mg, \red{we compare our results with three models, namely, the ENDF data evaluation, and the KDUQ and as well as WP potentials.} Comparisons with the available ENDF evaluations indicate excellent agreement for energies above 65 MeV. Predictions from KDUQ align closely with our results, while WP starts to deviate at higher energies as heavier masses are considered. For $^{32}$Mg, we compare our findings only with KDUQ, since ENDF evaluations are not available, and WP is not applicable for nuclei with a large neutron-to-proton ratio. As observed with the other magnesium isotopes, the KDUQ values for $^{32}$Mg fall slightly below our calculations for energies between 100 and 200 MeV.

In examining the proton reaction cross sections, we observe a spread of \hw~values for $^{26}$Mg and $^{28}$Mg similar to that of $^{24}$Mg, as depicted in Fig. \ref{fig:26allMg_neut}(b). The ENDF evaluations agree \red{slightly more} with our calculations for the $p+^{26}$Mg reaction, consistently exceeding the KDUQ predictions up to 150 MeV. However, in the case of $p+^{28}$Mg, the ENDF proton reaction cross sections abruptly drop to levels comparable to KDUQ. Notably, unlike the ENDF evaluations, all \red{KDUQ, WP and our model} predict an increase in proton reaction cross sections from $^{26}$Mg to $^{28}$Mg. This trend is also expected due to the slight increase in the charge radius of $^{28}$Mg compared to $^{26}$Mg \cite{Yordanov:2012zz}. Hence, this \red{may point} to a need for a revised ENDF evaluation for $^{28}$Mg. 

When assessing the proton reaction cross sections from the WP potential, we find that it agrees with our model for both $^{26}$Mg and $^{28}$Mg, unlike in the case of $^{24}$Mg. However, we note that the cross sections from WP increase by nearly 100 mb between $^{24}$Mg and $^{26}$Mg, and then again between $^{26}$Mg and $^{28}$Mg, while our model and KDUQ exhibit a much more gradual rise in cross sections throughout the isotopic chain. Comparisons of our model with KDUQ reveal a consistent difference of approximately 10-15\% between the centroids across all magnesium isotopes. 

\begin{figure}[h]
\centering
\includegraphics[width=7.5cm]{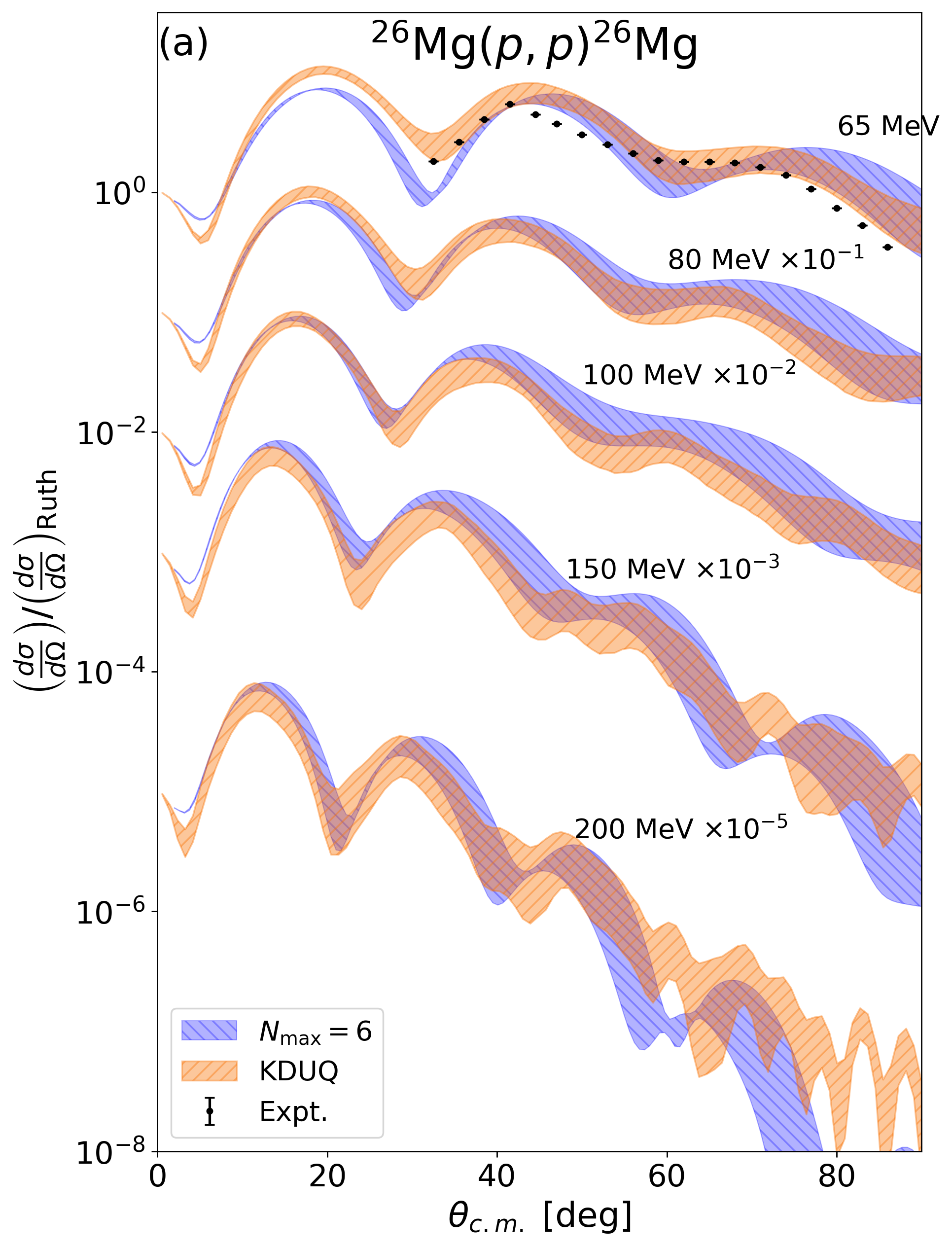}
\includegraphics[width=7.2cm]{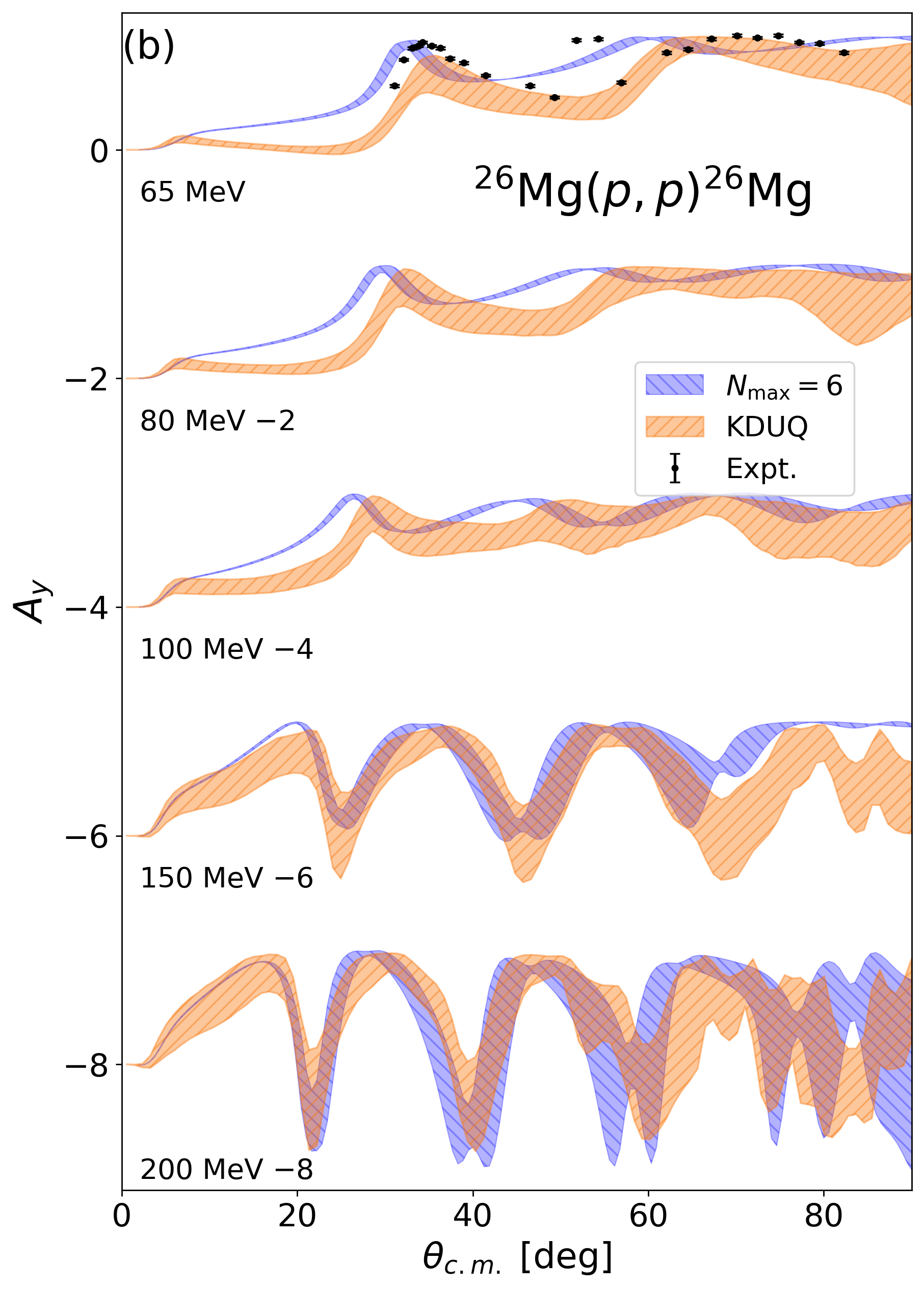}
\caption{The angular distribution of the differential cross section divided by the Rutherford cross section (a) and analyzing power (b) as a function of the c.m. scattering angle for elastic proton scattering from $^{26}$Mg from 65 to 200 MeV laboratory projectile energy. The band for the ``\Nmax=6'' calculations represents the variation of $\hw$ between 13 and 17 MeV. The cross sections are multiplied by the powers of 10 indicated at the energies listed in the figure, while the analyzing powers are offset by -2. The data at 65 MeV are from \cite{Nakamura:1981oja,HSakaguchi}.
}
\label{fig:26Mg_stacked}
\end{figure}

\begin{figure}[h]
\centering
\includegraphics[width=7.5cm]{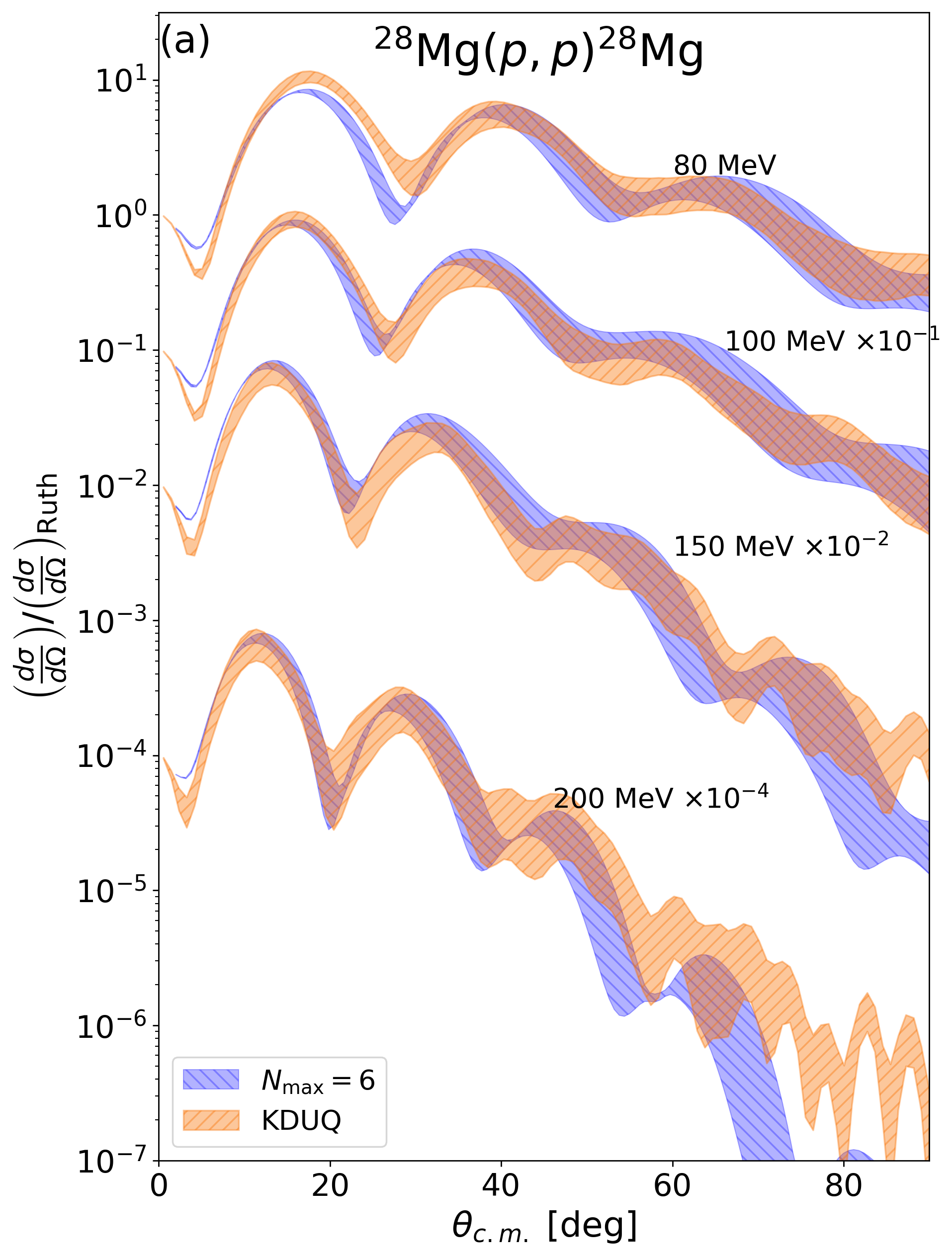}
\includegraphics[width=7.2cm]{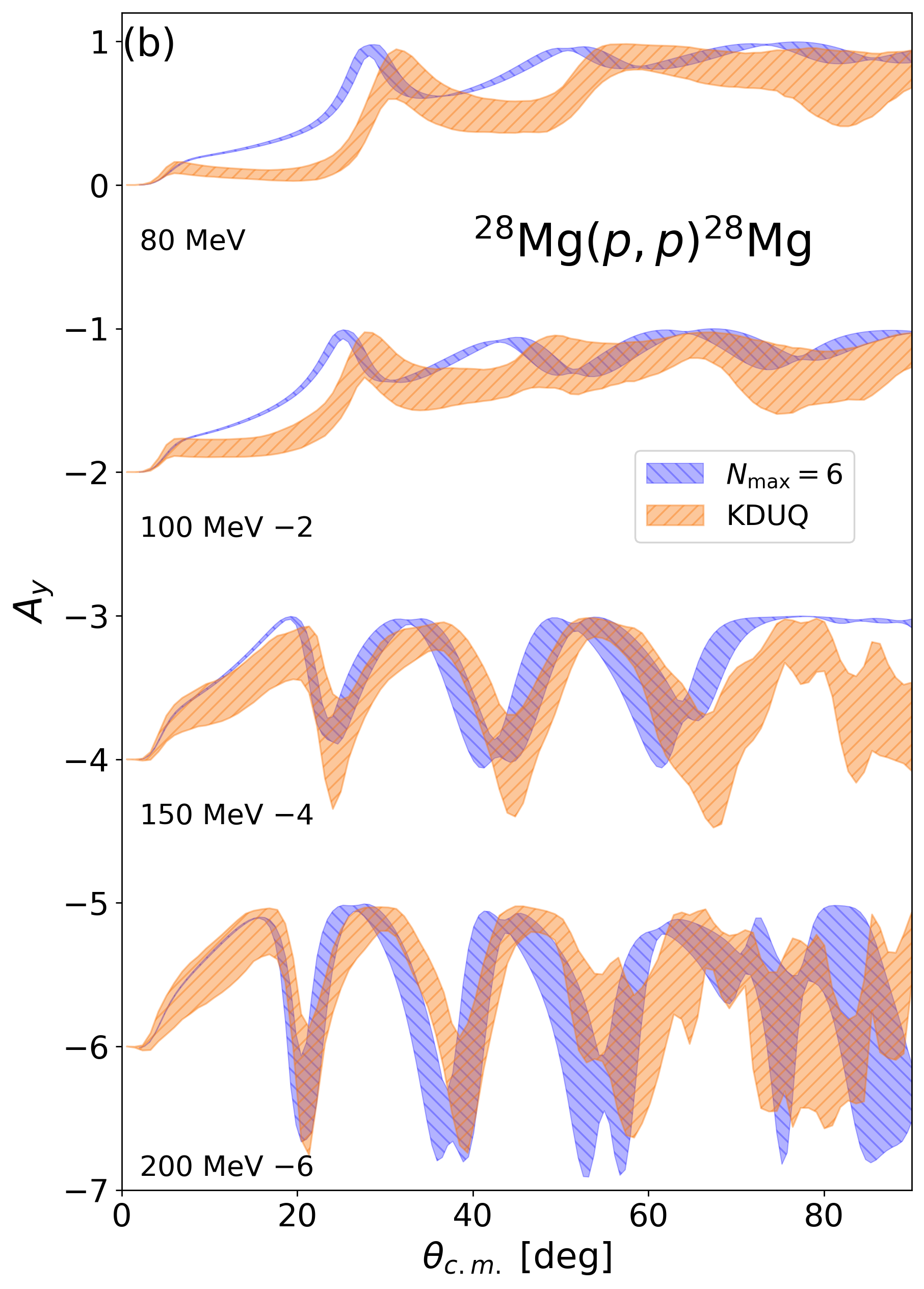}
\caption{The angular distribution of the differential cross section divided by the Rutherford cross section (a) and analyzing power (b) as a function of the c.m. scattering angle for elastic proton scattering from $^{28}$Mg from 80 to 200 MeV laboratory projectile energy. The band for the ``\Nmax=6'' calculations represents the variation of $\hw$ between 13 and 17 MeV. The cross sections are multiplied by the powers of 10 indicated at the energies listed in the figure, while the analyzing powers are offset by -2.}
\label{fig:28Mg_stacked}
\end{figure}

\begin{figure}[h]
\centering
\includegraphics[width=7.5cm]{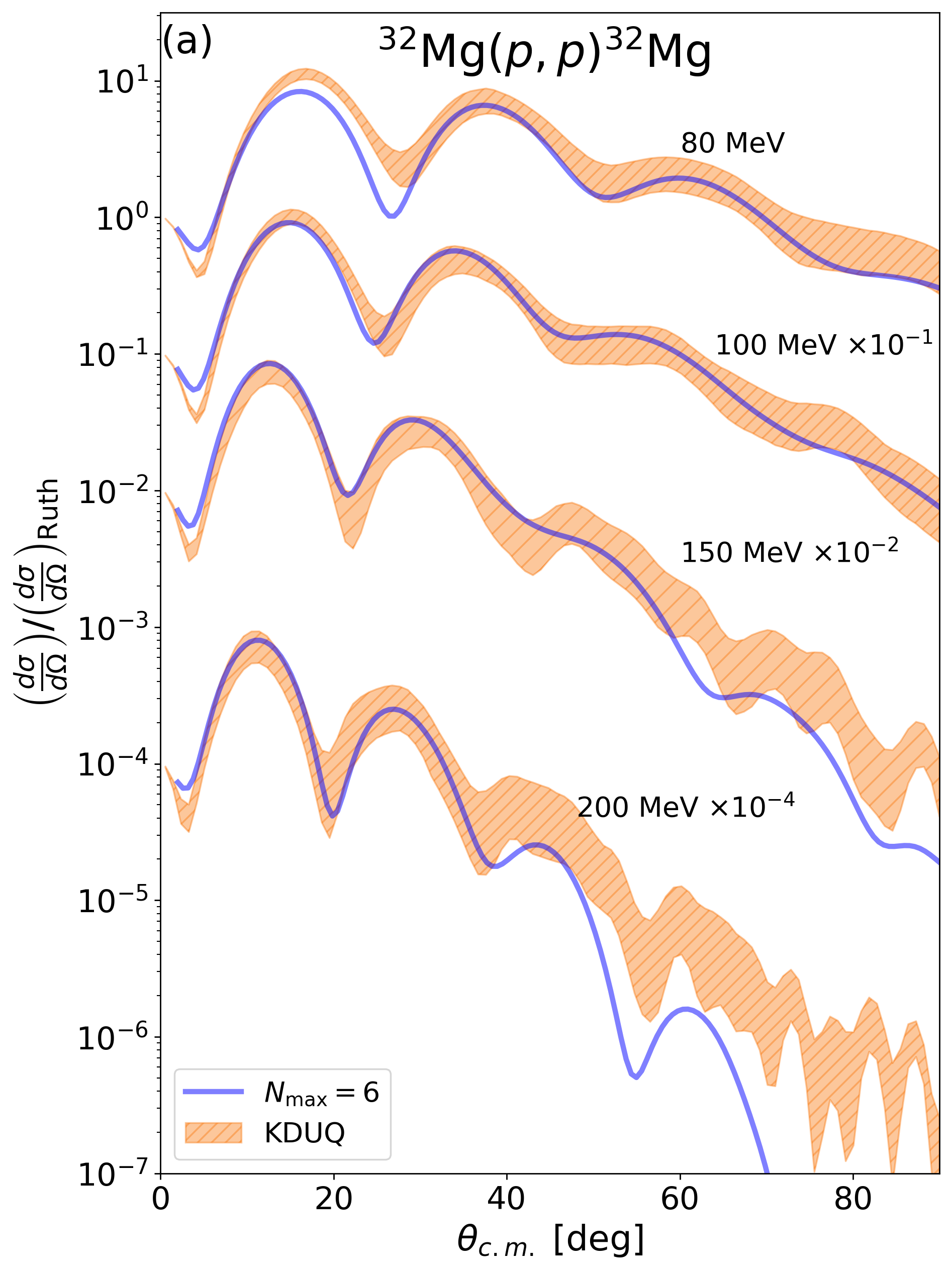}
\includegraphics[width=7.2cm]{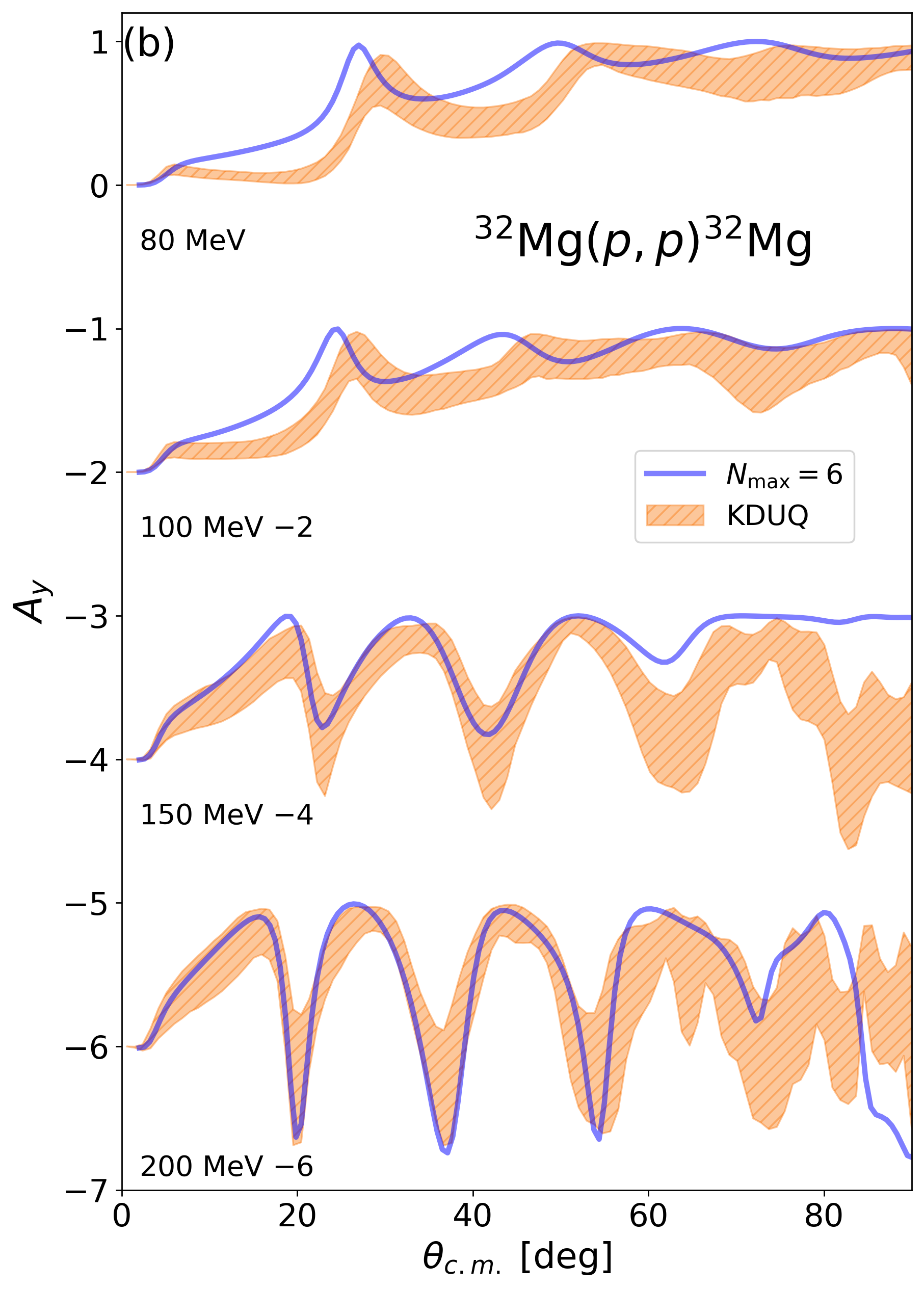}
\caption{The angular distribution of the differential cross section divided by the Rutherford cross section (a) and analyzing power (b) as a function of the c.m. scattering angle for elastic proton scattering from $^{32}$Mg from 80 to 200 MeV laboratory projectile kinetic energy. The \Nmax=6 calculations are done with $\hw$=15 MeV. The cross sections are multiplied by the powers of 10 indicated at the energies listed in the figure, while the analyzing powers are offset by -2.}
\label{fig:32Mg_stacked}
\end{figure}

To provide further insights into the proton-scattering observables for each isotope, we present the elastic-scattering cross sections and analyzing powers for various projectile energies in Figs. \ref{fig:26Mg_stacked}, \ref{fig:28Mg_stacked}, and \ref{fig:32Mg_stacked}. Our results are compared with calculations from KDUQ, which is anticipated to perform effectively near stability (e.g., for $^{26}$Mg). However, the extent of its performance in the more neutron-rich isotopes remains uncertain.

Examining $^{26}$Mg differential cross sections, we observe good agreement between our potential and KDUQ  within their respective uncertainties (Fig. \ref{fig:26Mg_stacked}(a)). Some deviations occur at higher angles (momentum transfer), where the cross sections from the {\it ab initio} leading-order potential slightly drop faster than those from KDUQ, indicating larger absorption. This increased absorption corresponds to the larger proton reaction cross sections illustrated in Fig. \ref{fig:26allMg_neut}(b) compared to KDUQ. Additionally, we note an overall good agreement between the two models for $A_y$. At a proton energy of 65 MeV, we compare our calculated $A_y$ to the available experimental data from Ref. \cite{Nakamura:1981oja}. The calculations align well with the first peak of the experimental data and exhibit only minor deviations at larger angles. 

Comparing the calculations of angular distributions for $^{28}$Mg to $^{26}$Mg, we notice slight shifts of the minima and maxima toward the smaller angles (Fig. \ref{fig:28Mg_stacked}). This phenomenon can be attributed to the marginally larger radius of $^{28}$Mg compared to $^{26}$Mg \cite{Yordanov:2012zz}. Overall, within the considered energy range, the behavior and the magnitude of the proton elastic differential cross sections and analyzing power for $^{28}$Mg are similar to $^{26}$Mg. The calculations using KDUQ are again in good agreement with our results for the forward angles, with small differences appearing at angles corresponding to larger momentum transfer.

Finally, we present the differential elastic cross sections and analyzing power of proton scattering on $^{32}$Mg for energies ranging from 80 to 200 MeV, as illustrated in Fig. \ref{fig:32Mg_stacked}. It is important to note that our calculations are for one model space and \hw~only due to the high computational cost to model $^{32}$Mg structure \emph{ab initio}. Thus, \red{a more complete \emph{ab initio} treatment of $^{32}$Mg would require more \hw~and model spaces.}
Given that $^{32}$Mg has a larger radius, its diffraction pattern is slightly more compressed and shifted toward smaller angles compared to the lighter magnesium isotopes. Calculations with KDUQ are in good agreement with our \red{ {\it ab initio} calculations} for the forward angles. 

Notably, the uncertainties in \red{the} KDUQ calculations do not increase significantly between the stable $^{24}$Mg and the neutron-rich $^{32}$Mg, as one might expect when extrapolating away from stability. This potentially highlights limitations in the linear dependence of potential well depths on $(N-Z)/A$ used in most global phenomenological optical potentials: while this simple form describes elastic nucleon scattering near stability well, and is reasonably constrained there, it is not guaranteed \emph{a priori} that the same systematic is valid out to the drip lines. Therefore, one would expect that KDUQ under-predicts uncertainties for isotopes at the asymmetry extremes, due to the inherent epistemic uncertainty of specifying a linear model.  Future phenomenological approaches could investigate higher orders of $(N-Z)/A$, which would naturally be less constrained by data near stability, leading to uncertainties that increase with extrapolation. Alternatively, approaches that attempt to infer the inherent model discrepancy of the optical potential along with its parameters (e.g., \cite{kennedy2001bayesian}) could also provide better uncertainty estimates away from stability. We expect that refining the model form in future phenomenological efforts will greatly benefit from microscopic approaches like ours.

On the whole, however, the comparison between our microscopic model and KDUQ \red{supports the use of KDUQ within the considered kinematic region} 
when extrapolated to the neutron-rich magnesium isotopes. Since $^{32}$Mg is of interest to several ongoing and planned experiments at rare-isotope beam facilities \cite{frib_experiments}, such \red{a consistency check} is vital, as KDUQ and its original version, the Koning-Delaroche potential, are commonly used for the analysis of experimental data. Moreover, our extracted \textit{ab initio} optical potentials can be generated on demand for any energies between 65 and 250 MeV and used as inputs in calculations of reaction observables beyond the ones discussed in this manuscript, such as knock-out and charge-exchange reactions.  

\section{Summary and Outlook}
\label{sec:summary}
In this work, we present the first calculations of {\it ab initio} nonlocal optical potentials for a chain of magnesium isotopes using the leading-order term of the spectator expansion of the multiple scattering theory. 
Based on the SA-NCSM, we are able to reliably describe the structure of these isotopes and provide the off-shell, translationally invariant scalar and spin-projected densities serving as input for the construction of the {\it ab initio} leading-order optical potentials, in which structure and reaction input is treated on equal footing.
 In particular, we studied magnesium isotopes from the mid-shell N=Z~$^{24}$Mg to the N=20 island of inversion $^{32}$Mg. All isotopes exhibit strong deformation and collectivity, hence pose a challenge to many \emph{ab initio} methods, but can be well described by the symmetry-adapted basis of the SA-NCSM. 
 Lacking free parameters, our {\it ab initio} optical potential has strong predictive power, enabling rigorous calculations for unstable systems such as $^{32}$Mg, for which experimental data are scarce. 

First, we demonstrated the quality of the {\it ab initio} optical potential in leading order in the spectator expansion by comparing calculated reaction observables to experimental measurements for $^{24}$Mg. Our \emph{ab initio} predictions for the total cross section for neutron scattering from  $^{24}$Mg between 100 and 250 MeV reproduce the experimental data perfectly. Minor deviations occur below 80 MeV, where rescattering effects (i.e., higher-order spectator terms) may become visible. 
We showed good agreement with experimental proton differential cross sections and analyzing powers from 65 to 250 MeV, with improved agreement at higher energies. Here as well, the low-energy discrepancies likely reflect omitted rescattering, whereas at the highest energy (250 MeV), the deviations can most likely be attributed to using the chiral \NNLOopt  potential beyond its intended range of applicability.

\red{For a comparison with the phenomenological optical potentials we mainly focused on the KDUQ potential. For the total neutron cross sections and the proton reaction cross sections, we included the WP phenomenological potential as well as the ENDF evaluations. Our comparisons reveal} that KDUQ and WP systematically underpredict proton reaction cross sections relative to our \emph{ab initio} calculation and ENDF for $^{24}$Mg. Moreover, the partial-wave contributions to the reaction cross section indicate the dominant contributions arise at the same angular momentum, although
KDUQ yields significantly smaller partial contributions compared to our calculation. 

Then we extend our calculations to predict scattering observables for $^{26}$Mg, $^{28}$Mg, and $^{32}$Mg, using ground state densities from SA-NCSM. We observe a systematic rise in neutron total cross sections from $^{24}$Mg to the heavier isotopes that agree perfectly with the ENDF evaluations above $\sim65$ MeV for  $^{26}$Mg and $^{28}$Mg. Calculations with KDUQ generally track our results, with very minor deviations at $\sim 65$ MeV.  For proton reaction cross sections, our potential and KDUQ show a modest, steady increase across the isotopic chain (consistent with a small growth in charge radius), whereas the WP potential predicts much larger jumps between isotopes. However, KDUQ calculations of proton reaction cross sections remain $\sim 10-15$\% lower than our results across the magnesium chain, a feature that we link to the systematically larger imaginary part of the microscopic potential. 

For a more detailed study of the proton elastic scattering observables as we move up in the magnesium isotopic chain, we have also presented angular distributions of elastic scattering cross sections and analyzing powers for $^{26}$Mg, $^{28}$Mg, and $^{32}$Mg. For $^{26}$Mg, the calculated differential cross sections and $A_y$ agree well with KDUQ and with available 65 MeV data at forward angles, with differences appearing at larger angles and momentum transfers.  Moving to $^{28}$Mg and $^{32}$Mg, diffraction features shift to smaller angles in line with slightly larger radii. KDUQ maintains good forward-angle agreement even for neutron-rich $^{32}$Mg, supporting its continued use in experimental analyses.

With this work, we are able to show a constructive interplay between {\it ab initio} optical potentials that predict elastic scattering observables for exotic nuclei close to the dripline 
and phenomenological optical potentials for which parameters are fitted along the valley of stability and which are extrapolated towards nuclei at the dripline. 
Although KDUQ, \red{our specific choice of phenomenological optical potential,} is extrapolated with uncertainty quantifications, our work should give users of this specific potential a higher level of confidence when using it, e.g., in reaction codes like, e.g., TALYS \cite{koning2023talys} when analyzing reactions with nuclear masses around A$\simeq$30 and the energy regime we covered in our study.
\red{Our {\it ab initio} calculations of the scattering observables together with the KDUQ results which are included in the supplemental material can be used to test other phenomenological optical potentials and their extrapolation towards the dripline for the Magnesium isotopes.} 

One may speculate that an extrapolation towards the proton dripline may behave similarly. This could be \red{investigated in} a further study. In general, computing optical potentials {\it ab initio}
in an energy regime greater than $\approx$ 60~MeV requires many-body methods to compute off-shell nuclear density matrices, which are input to folding calculations when used within a multiple scattering theory to obtain the optical potential. This is usually computationally involved. 
If our comparison between {\it ab initio} and extrapolated phenomenological optical potentials toward the dripline can be repeated in different mass regimes of the nuclear chart, it can elevate the confidence in the use of phenomenological optical potentials in a variety of reaction codes.


\ack{The authors are grateful to Robert Baker and Matthew Burrows for sharing their useful scripts. We thank Filomena Nunes, Remco Zegers, Kristina Launey, Gregory Potel, Chlo\"e Hebborn and Stephen Weppner for enlightening discussions. CE appreciates the support of the FRIB nuclear theory group during her sabbatical stay at FRIB.}

\funding{

This material is based upon work supported by the U.S. Department of Energy, Office of Science, Office of Nuclear Physics, under the FRIB Theory Alliance award DE-SC0013617, and contract DE-FG02-93-ER40756 with Ohio University.

The work of KB was supported by the National Science Foundation CSSI program under award number OAC-2004601 (BAND Collaboration).

The numerical computations benefited from computational resources provided by the National Energy Research Scientific Computing Center (NERSC), a U.S.~Department of Energy Office of Science User Facility using NERSC awards NP-ERCAP0033451 and NP-ERCAP0035912. This work also benefited from high-performance computational resources provided by the Frontera computing project at the Texas Advanced Computing Center (TACC), made possible by National Science Foundation award OAC-1818253, as well as in part through computational resources and services provided by the Institute for Cyber-Enabled Research at Michigan State University. }


\data{All data that support the findings of this study are included within the article (and any supplementary files).}

\section*{Appendix A: Nuclear one-body densities from SA-NCSM}

\begin{figure}[ht]
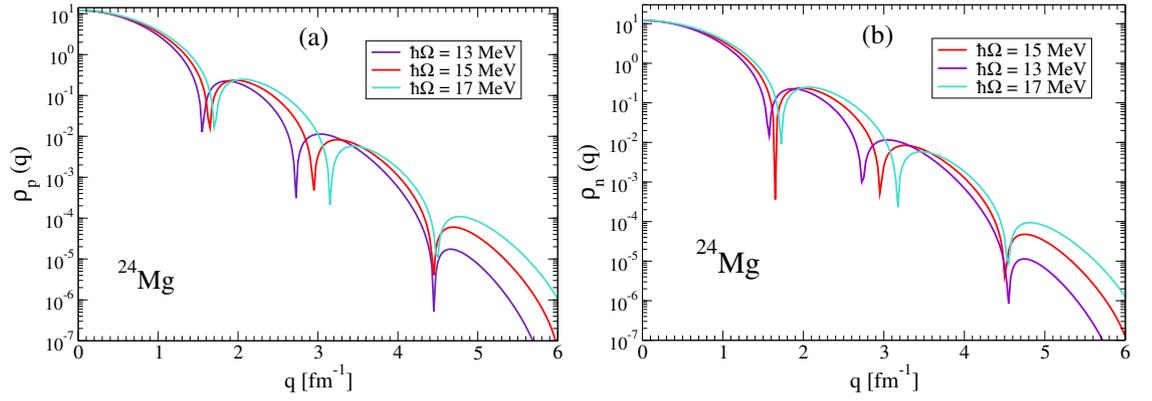

\centering
\includegraphics[width=7.3cm]{figs-appendix/diagonal_mg24p-hw-log.eps}
\includegraphics[width=7.3cm]{figs-appendix/diagonal_mg24n-hw-log.eps}
\caption{The diagonal one-body (a) proton and (b) neutron densities vs. momentum transfer for $^{24}$Mg for \Nmax=6 and \hw=13, 15, and 17 MeV.}
\label{fig:24Mg_dens_hw}
\end{figure}
The primary structure input in the leading-order term of the spectator expansion of multiple-scattering theory is the one-body nucleon density. Fig. \ref{fig:24Mg_dens_hw} presents the proton and neutron densities vs. the momentum transfer ($q$) in $^{24}$Mg for \hw=13, 15, and 17 MeV and \Nmax=6 calculated using SA-NCSM. Small values of $q$ would correspond to large values of $r$ in the coordinate space. The weak \hw~dependence of the density at small values of $q$ results in the narrow \hw~bands at forward angles in the elastic scattering calculations (Fig. \ref{fig:24Mg_stacked}). 
\begin{figure}[h]
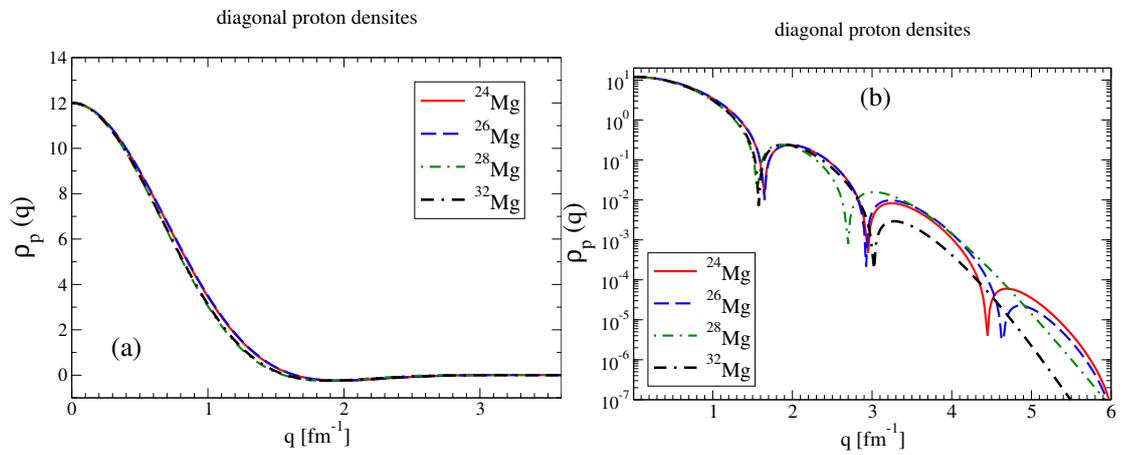

\centering
\includegraphics[width=7.3cm]{figs-appendix/diagonal-hw15-Nm6-proton-all.eps}
\includegraphics[width=7.3cm]{figs-appendix/diagonal-hw15-Nm6-proton-all-log.eps}
\caption{The diagonal one-body proton and densities  (a) linear and (b) logarithmic scale vs. momentum transfer in for $^{24}$Mg, $^{26}$Mg, $^{28}$Mg and $^{32}$Mg for \Nmax=6 and \hw=15 MeV.}
\label{fig:Mg_p_dens_all}
\end{figure}
\begin{figure}[h]
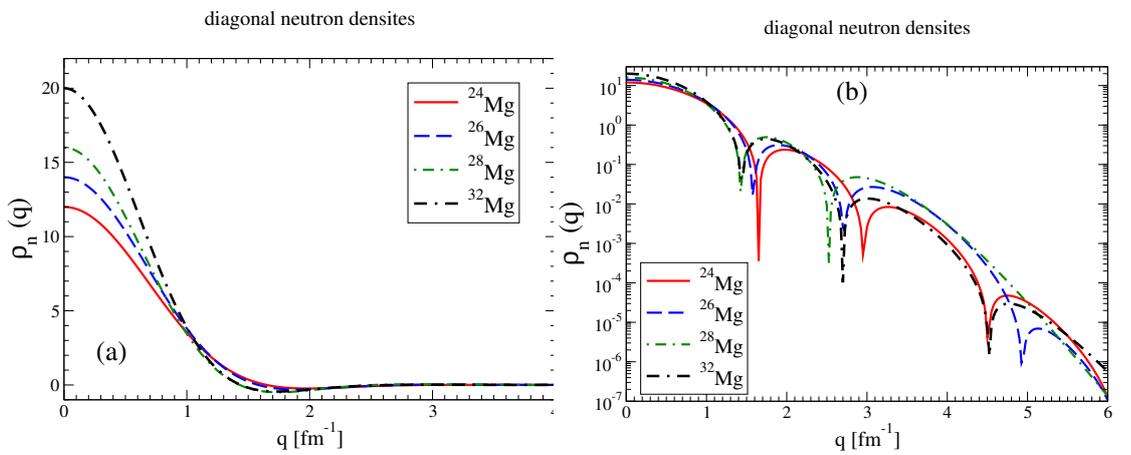

\centering
\includegraphics[width=7.3cm]{figs-appendix/diagonal-hw15-Nm6-neutron-all.eps}
\includegraphics[width=7.3cm]{figs-appendix/diagonal-hw15-Nm6-neutron-all-log.eps}
\caption{The diagonal one-body neutron and densities  (a) linear and (b) logarithmic scale vs. momentum transfer in for $^{24}$Mg, $^{26}$Mg, $^{28}$Mg and $^{32}$Mg for \Nmax=6 and \hw=15 MeV.}
\label{fig:Mg_n_dens_all}
\end{figure}
The proton and neutron one-body densities for \hw=15  MeV and \Nmax=6 for all four isotopes discussed in this manuscript are presented in Figs. \ref{fig:Mg_p_dens_all} and \ref{fig:Mg_n_dens_all} both in linear and logarithmic scale. In a linear scale, one can clearly see the number of particles in a given isotope at $q=0$. We again note that the densities in momentum space that have higher peaks at $q=0$ and drop faster at a certain range (e.g., $^{32}$Mg vs $^{24}$Mg neutron densities up to 1.5 fm$^{-1}$) would correspond to lower peak at $r=0$ and slower decrease with $r$ (longer tail) in the coordinate space.

\red{We also present the point-proton, neutron and charge radii from SA-NCSM calculations. For $^{24}$Mg, $^{26}$Mg and $^{28}$Mg, we also present charge radii extrapolated to infinite model space and compared to the experiment. Extrapolations are performed using Richardson's method \cite{Richardson1910}. Charge radii are obtained from the point proton radii using $r_c=\sqrt{r_p^2+r_{op}^2+(N/Z)r_{on}^2+3/(4M_p^2)}$, where $r_p$ is the point proton rms radius calculated in SA-NCSM, $r_{op}=0.84075(64)$ fm is the mean-square radius of the proton from \cite{MohrNTT2025}, $3/(4M_p^2)=0.033$ fm$^2$ is the Darwin-Foldy term \cite{particle2020review}, $r_{on}^2=-0.1161(22)$ is the neutron mean square charge radius \cite{particle2020review}. All extrapolated charge radii are in agreement with the experiment given the theoretical and experimental uncertainties. }
\begin{table}[h]
    \centering
    \begin{tabular}{|c|c|l|l|c|}\hline
         $\hbar\Omega$ (MeV)&  \Nmax& RMS-p&RMS-n& Charge radius\\ \hline \hline
         13& 6 & 2.898 & 2.877 & 3.004 \\
         15&   4 & 2.679&2.662& 2.793\\  
           &  6 & 2.737 & 2.718 & 2.849 \\
         &   8&    2.765&2.744& 2.876\\
         17 & 6 & 2.613 & 2.596 & 2.742 \\
         Extrap.&  &  &  & 2.94(14) \\
         \hline
    \end{tabular}
    \caption{The calculated point proton (RMS-p), point-neutron (RMS-n) and charge rms-radii in fm for the $^{24}$Mg densities used in the scattering calculations along with charge radius extrapolation to the infinite \Nmax. The experimental charge radius for $^{24}$Mg is 3.057(5) fm \cite{Yordanov:2012zz}. }
    \label{tab:24Mgrms-radii}
\end{table}

\begin{table}[h]
\centering
\begin{tabular}{|c|c|l|l|c|}\hline
         $\hbar\Omega$ (MeV)&  \Nmax& RMS-p&RMS-n& Charge radius\\ \hline \hline
         13& 6 & 2.871 & 2.938 & 2.974 \\
         15& 2 & 2.668 & 2.722 & 2.779\\
            &4 & 2.688 & 2.750& 2.798\\
           &  6 & 2.711 & 2.774 & 2.820 \\
         17 & 6 & 2.582 & 2.638 & 2.697 \\
         Extrap.&  &  &  &  2.90(13) \\
         \hline
\end{tabular}
\caption{Same as Table \ref{tab:24Mgrms-radii} but for $^{26}$Mg. The experimental charge radius for $^{26}$Mg is 3.034(3)~fm~\cite{Yordanov:2012zz}. 
}
\label{tab:rms-radii-26}
\end{table}

{\begin{table}[h]
\centering
\begin{tabular}{|c|c|l|l|c|}\hline
         $\hbar\Omega$ (MeV)&  \Nmax& RMS-p&RMS-n& Charge radius\\ \hline \hline
         13& 6 & 2.999 & 3.121 &  3.095 \\
         15& 2 & 2.763 & 2.873 & 2.867 \\
           & 4 & 2.802 & 2.913 & 2.905 \\
           &  6 & 2.831 & 2.942 & 2.933 \\
         17 & 6 & 2.694 & 2.694 &  2.801\\
            Extrap.&  &  &  & 3.02(13) \\
         \hline
\end{tabular}
\caption{Same as Table \ref{tab:24Mgrms-radii} but for $^{28}$Mg. The experimental charge radius for $^{28}$Mg is 3.0695(14)~fm~\cite{Yordanov:2012zz}. 
}
\label{tab:rms-radii-28}
\end{table}

\begin{table}[h]
\centering
\begin{tabular}{|c|c|l|l|c|}\hline
         $\hbar\Omega$ (MeV)&  Nmax& RMS-p&RMS-n& Charge radius\\ \hline \hline
         15  &  6 & 2.802 & 3.052 & 2.909 \\
         \hline
\end{tabular}
\caption{The calculated rms-radii in fm for the $^{32}$Mg densities used in the scattering calculations. 
}
\label{tab:rms-radii-32}
\end{table}

\section*{Appendix B: Volume integrals of optical potentials}

For completeness, in Fig. \ref{fig:Mg_volint} we present the volume integrals for proton scattering on the $^{26}$Mg, $^{28}$Mg, and $^{32}$Mg potentials from KDUQ and our microscopic calculations as a function of the proton laboratory energy. For both methods, the patterns are similar to the ones seen in $^{24}$Mg (Fig. \ref{fig:mg24_volint}). The bands on our microscopic potential results are only due to the variation of the harmonic oscillator parameter \hw~between 13 and 17 MeV, while the  bands on KDUQ correspond to the sampling of all $\sim$40 parameters. 
\begin{figure}[b]
\centering
\includegraphics[width=15.2cm]{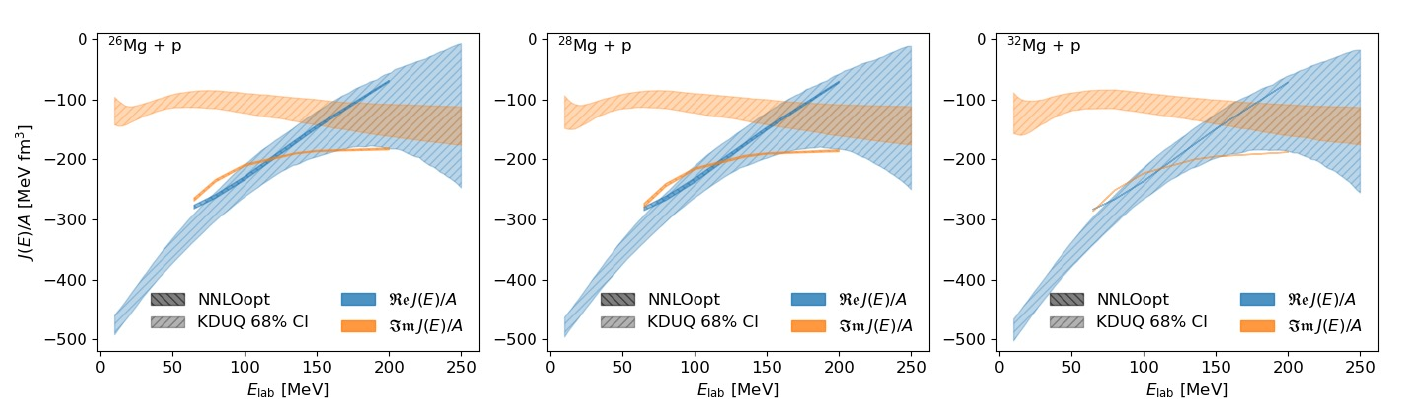}
\caption{Volume integrals per nucleon of optical potentials for proton + $^{26}$Mg, $^{28}$Mg and $^{32}$Mg for KDUQ and the microscopic potential (denoted ``NNLOopt'') in \Nmax=6. The bands on the microscopic potential correspond to \hw~variance between 13 and 17 MeV except for $^{32}$Mg, where the lines correspond to the \hw=15 MeV. }
\label{fig:Mg_volint}
\end{figure}

\clearpage
 \section*{Appendix C: Selected observables calculated with the Whitehead-Lim-Holt optical potential}

For comparison, we also present cross sections calculated using an optical potential derived from chiral $NN$ interactions but within a nuclear matter framework from Ref. \cite{WhiteheadLH2019,WhiteheadLH2020,WhiteheadLH2021} introduced in Sec. \ref{sec:intro} (referred to as WLH potential). 
We show in Fig.~\ref{fig:Mg-WLH} calculations for proton scattering on $^{24}$Mg at 135~MeV and on $^{32}$Mg at 100~MeV with our model, WLH, and KDUQ potentials. We note that the WLH potential is not derived from the same \NNLOopt~chiral interaction parametrization that was used in our spectator expansion calculations.  

\begin{figure}[h]
    \centering
    \includegraphics[width=0.45\linewidth]{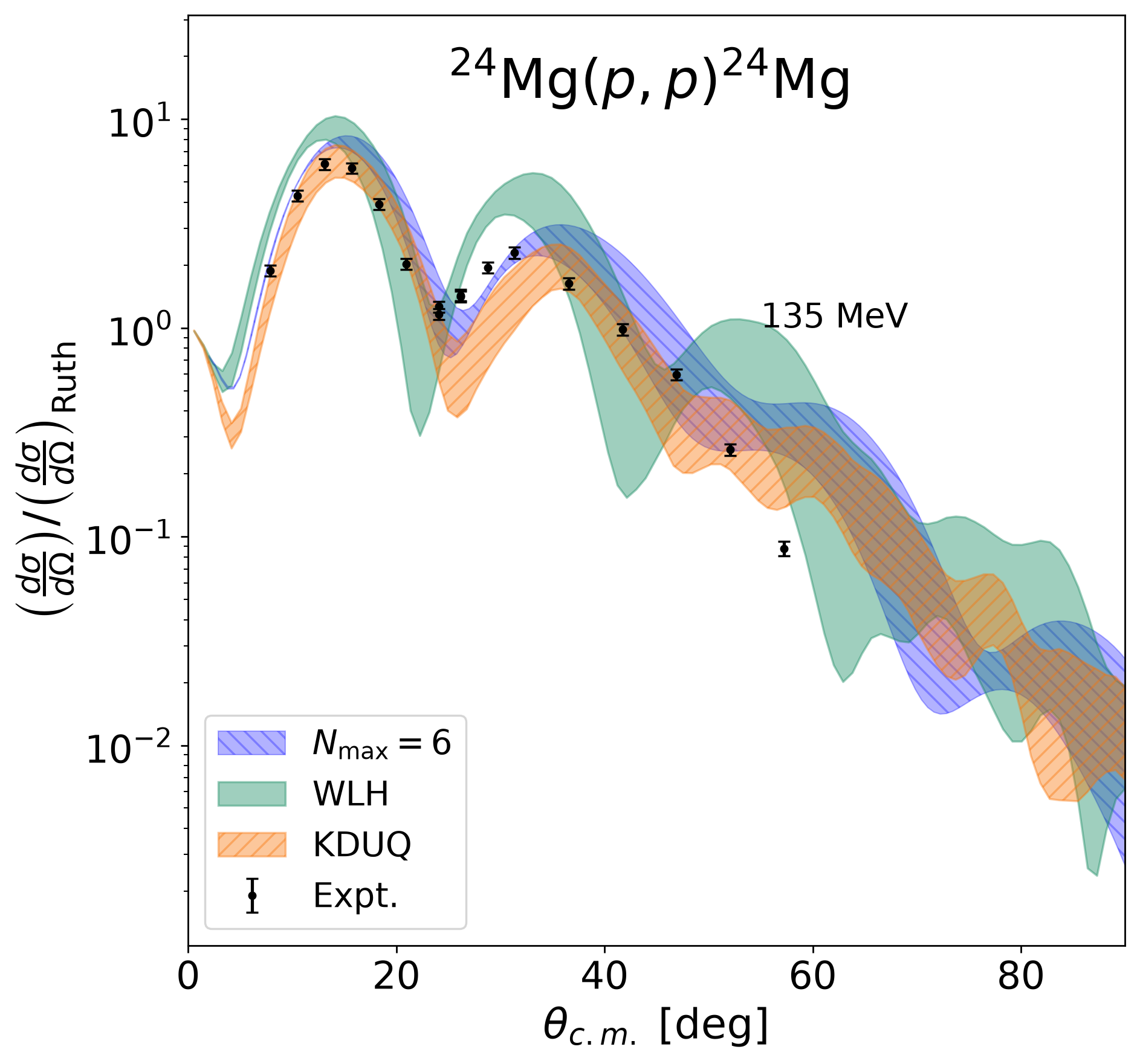}
    \includegraphics[width=0.45\linewidth]{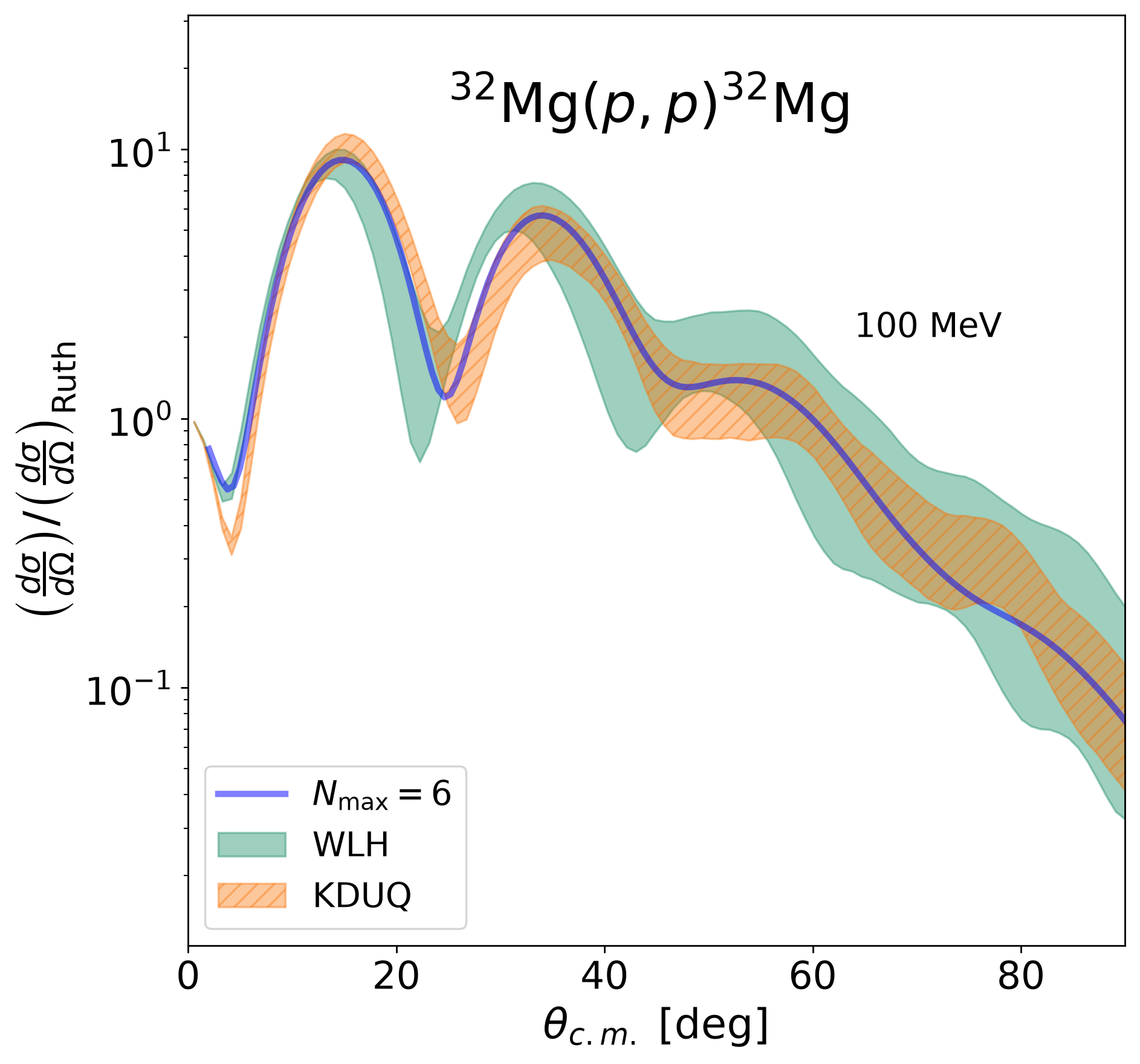}
    \caption{The differential cross sections for elastic scattering from $^{24}$Mg at 135~MeV laboratory kinetic energy and from $^{32}$Mg at 100~MeV, calculated at leading-order in the spectator expansion using SA-NCSM densities, and WLH and KDUQ potentials. The experimental data for $^{24}$Mg($p,p$) are from Ref.~\cite{Schwandt:1982py}.}
    \label{fig:Mg-WLH}
\end{figure}


\bibliographystyle{unsrt}
\bibliography{Mgisotopes}

\end{document}